\documentclass[%
 reprint,
 superscriptaddress,
 nofootinbib,
 twocolumn, 
 amsmath,amssymb,
 aps,
prb,
]{revtex4-2}

\usepackage[colorlinks = true,
            linkcolor = blue,
            urlcolor  = red,
            citecolor = red,
            anchorcolor = blue]{hyperref}
\usepackage[T1]{fontenc}
\usepackage[utf8]{inputenc}
\usepackage{todonotes}
\usepackage{graphicx}
\usepackage{dcolumn}
\usepackage{color}
\usepackage{amsmath}
\usepackage{braket}
\usepackage{appendix}
\usepackage{MnSymbol}
\usepackage{orcidlink}
\usepackage{glossaries}
\newacronym{CDW}{CDW}{charge-density-wave}
\newacronym{RIXS}{RIXS}{resonant inelastic x-ray scattering}
\newacronym{DMRG}{DMRG}{density matrix renormalization group}
\newacronym{DQMC}{DQMC}{determinant quantum Monte Carlo}
\newacronym{QMC}{QMC}{quantum Monte Carlo}
\newacronym{SSH}{SSH}{Su-Schrieffer-Heeger}
\newacronym{oSSH}{oSSH}{optical Su-Schrieffer-Heeger}
\newacronym{2D}{2D}{two-dimensional}
\newacronym{1D}{1D}{one-dimensional}
\newacronym{FS}{FS}{Fermi surface}
\newacronym{eph}{$e$-ph}{electron-phonon}
\newacronym{HMC}{HMC}{hybrid Monte Carlo}
\newacronym{KVB}{KVB}{Kekul{\'e} Valence Bond}
\newacronym{KVBS}{KVBS}{Kekul{\'e} Valence Bond Solid}
\newacronym{GS}{GS}{ground state}
\newacronym{QCP}{QCP}{quantum critical point}
\newacronym{FSS}{FSS}{finite-size scaling}
\newacronym{AFM}{AFM}{antiferromagentic}
\newacronym{SC}{SC}{superconducting}
\newacronym{HSSH}{HSSH}{Hubbard-SSH}
\newacronym{SM}{SM}{supplementary materials}

\begin{document}

\preprint{}
\title{Spectral signatures of residual electron pairing in the extended-{H}ubbard--{S}u-{S}chrieffer-{H}eeger model}

\author{Debshikha Banerjee\orcidlink{0009-0001-2925-9724}}
\affiliation{Department of Physics and Astronomy, The University of Tennessee, Knoxville, Tennessee 37996, USA}
\affiliation{Institute for Advanced Materials and Manufacturing, University of Tennessee, Knoxville, Tennessee 37996, USA\looseness=-1}

\author{Alberto Nocera\orcidlink{0000-0001-9722-6388}}
\affiliation{Department of Physics Astronomy, University of British Columbia, Vancouver, British Columbia, Canada V6T 1Z1}
\affiliation{Stewart Blusson Quantum Matter Institute, University of British Columbia, Vancouver, British Columbia, Canada
V6T 1Z4}

\author{George A. Sawatzky\orcidlink{0000-0003-1265-2770}}
\affiliation{Department of Physics Astronomy, University of British Columbia, Vancouver, British Columbia, Canada V6T 1Z1}
\affiliation{Stewart Blusson Quantum Matter Institute, University of British Columbia, Vancouver, British Columbia, Canada
V6T 1Z4}

\author{Mona Berciu\orcidlink{0000-0002-6736-1893
}}
\affiliation{Department of Physics Astronomy, University of British Columbia, Vancouver, British Columbia, Canada V6T 1Z1}
\affiliation{Stewart Blusson Quantum Matter Institute, University of British Columbia, Vancouver, British Columbia, Canada
V6T 1Z4}

\author{Steven Johnston\orcidlink{0000-0002-2343-0113}}
\affiliation{Department of Physics and Astronomy, The University of Tennessee, Knoxville, Tennessee 37996, USA}
\affiliation{Institute for Advanced Materials and Manufacturing, University of Tennessee, Knoxville, Tennessee 37996, USA\looseness=-1}

\date{\today}

\begin{abstract}
We study the electron addition spectrum of the one-dimensional extended Hubbard-Su-Schrieffer-Heeger (HSSH) model in the dilute limit using the density matrix renormalization group method. In addition to the expected renormalization to the band structure, we find that the electron-phonon ($e$-ph) interaction produces an anomalous spectral feature when electrons are added in the singlet channel but which is absent in the triplet channel. By comparing these results with those obtained from perturbation theory in the antiadiabatic limit, we demonstrate that this anomalous feature is a remnant of the strong electron-electron interaction mediated by the SSH coupling previously derived in the two-particle limit. By studying the evolution of this feature as a function of doping, we track the fate of this attraction to higher carrier concentrations and provide predictions for the spectral features to help guide future searches for strong $e$-ph mediated pairing. 
\end{abstract}

\maketitle
 

\noindent{\bf Introduction} --- Understanding the fundamental limitations of phonon-mediated electron pairing is a long-standing problem in condensed matter physics. This question has important implications for determining the maximum superconducting transition temperature $T_\mathrm{c}$ in conventional superconductors and is a topic of active research~\cite{Achroft1968metallic, Cohen1972comments, Allen1983theory, chakraverty1998experimental, Kulic1994influence, Kulic2000interplay, Moussa2006twobounds, Gorkov2018colloquium, Esterlis2018bound,  Esterlis2018breakdown, Esterlis2018bound, Hazra2019bounds, Esterlis2019pseudogap, Hofmann2022heuristic, Pickett2023colloquium, Nosarzewski2021superconductivity, Bradley2021superconductivity, zhang2023bipolaronic, semenok2024fundamental, trachenko2025upper}. Eliashberg theory~\cite{Eliashberg} predicts that $T_\mathrm{c}$ grows without bound as a function of the dimensionless \gls*{eph} coupling $\lambda$ with $ T_\mathrm{c} \propto \sqrt{\lambda}$ in the strong coupling limit~\cite{Allen1975transition}, provided one can avoid competing \gls*{CDW} or structural instabilities or the formation of heavy (bi)polarons~\cite{Cohen1972comments, chakraverty1998experimental, Esterlis2018bound} that will ultimately suppress superconductivity. Indeed, several recent \gls*{QMC} studies of the Holstein model have observed the formation of (bi)polarons, which are prone to local ordering that ultimately suppresses superconductivity~\cite{Esterlis2018breakdown, Esterlis2019pseudogap, Nosarzewski2021superconductivity, Bradley2021superconductivity, issa2025learningconfusionphasediagram}. 

\gls*{SSH}- or Peierls-like interactions~\cite{Barisic1970tightbinding, Su1979solitons} have attracted considerable interest in this context. These interactions arise in systems where the atomic motion modulates the electronic hopping integrals rather than the on-site energies (as in the Holstein~\cite{Holstein1959studiesI, Holstein1959studiesII} or Fr{\"o}hlich~\cite{Froelich1952} models). Theoretical studies in the dilute limit have shown that \gls*{SSH} interactions mediate strong attractive $e$-$e$ interactions through pair hopping of on-site and nearest neighbor singlets~\cite{sous2018light}. Physically, this attraction appears when one particle hops by emitting a phonon while the other electron follows by absorbing it~\cite{Supplement,sous2018light}. 
This process leads to the formation of light polarons~\cite{Marchand10} and stable, mobile bipolarons~\cite{sous2018light} that can condense into a high-T$_\mathrm{c}$ superconductor~\cite{zhang2023bipolaronic}. The pair hopping processes can also give rise to novel forms of magnetism in the absence of a Coulomb interaction, including antiferromagnetism~\cite{Goetz2024phases, Cai2021antiferromagnetism, Casebolt2024magnetic} and quantum spin liquid behavior~\cite{cai2024quantum}. 

Recent \gls*{DQMC} studies have concluded that the details of the polaron mass, phonon-mediated magnetism, and superconducting correlations in \gls*{SSH} models depend on whether the phonon modes are placed on the lattice's bonds or sites (the so-called bond vs. optical/acoustic \gls*{SSH} models, respectively)~\cite{MalkarugeCosta2023comparative, TanjaroonLy2023comparative, TanjaroonLy2025antiferromagnetic}. In particular, the magnetism and robust electron pairing appear to be reduced for site phonons at finite carrier concentrations~\cite{TanjaroonLy2023comparative, TanjaroonLy2025antiferromagnetic}. \Gls*{DQMC} have also recently reported enhanced $T_\mathrm{c}$'s in simulations of the bond-\gls*{SSH} model near half-filling and in the antiadiabatic limit~\cite{cai2023hightemperature} but attributed it to a different mechanism. In light of these differences, studying the effective interactions mediated in different \gls*{SSH} models is necessary to test these theories further. 

In this letter, we examine the electron addition spectra of a \gls*{1D} extended Hubbard-\gls*{SSH} chain using \gls*{DMRG}. Our results reveal the presence of a strong resonance feature in the spectra that we attribute to the strong residual phonon-mediated attraction between electrons. Tracking this feature as a function of carrier concentration, we further demonstrate that it persists in finite carrier concentrations. Our results suggest that the strong phonon-mediated attraction persists in the optical \gls*{SSH} model beyond the dilute limit and identify a specific spectroscopic signature of this interaction that can guide experiments. \\


\noindent{\bf Model \& Methods} --- We study the optical variant of the \gls*{1D} extended-\gls*{HSSH} Hamiltonian ($\hbar=1$)~\cite{Capone1997small, MalkarugeCosta2023comparative}
\begin{align}\nonumber
    H&=-t \sum_{i,\sigma} \left[
    c^\dagger_{i,\sigma}c^{\phantom\dagger}_{i+1,\sigma} + \text{H.c.}\right] + U \sum_{i}\hat{n}_{i,\uparrow}\hat{n}_{i,\downarrow} + V\sum_{i} \hat{n}_i \hat{n}_{i+1}\\
    &+ \Omega \sum_{i}(b^\dagger_i b^{\phantom\dagger}_i+\tfrac{1}{2}) 
    + 
   g\sum_{i,\sigma} \left[c^\dagger_{i,\sigma}c^{\phantom\dagger}_{i+1,\sigma}\left(\hat{x}_i - \hat{x}_{i+1}\right) + \text{H.c.}\right].\label{eq:Hamiltonian_HSSH}
\end{align}
Here $c^\dagger_{i,\sigma}$ ($c^{\phantom\dagger}_{i,\sigma}$) creates (annihilates) a spin-$\sigma$ electron on lattice site $i$, $b^\dagger_i$ ($b^{\phantom\dagger}_{i}$) creates (annihilates) a phonon mode with energy $\Omega$ at site $i$, $\hat{x}_i = (b^\dagger_i + b^{\phantom\dagger}_i)= \sqrt{2M\Omega}\hat{X}_i $ is proportional to the displacement operator $\hat{X}_i$ for the $i$\textsuperscript{th} atom, $\hat{n}_i = \sum_{\sigma} \hat{n}_{i,\sigma}$ is the local electron density operator, $t$ is the nearest-neighbor hopping, $U$ and $V$ are the on-site and nearest-neighbor Hubbard repulsions, respectively, and $g$ is the \gls*{eph} coupling strength. Throughout this work, we set $M = t = 1$ and fix $\Omega=2t$, while varying $g$, $U$, and $V$. We have also explicitly confirmed that all of our presented results were obtained within the physical regime of the \gls*{SSH} model, where the sign of the effective hopping has not been inverted~\cite{Banerjee2023groundstate}. 

\begin{figure}[t]
    \centering
    \includegraphics[width=\columnwidth]{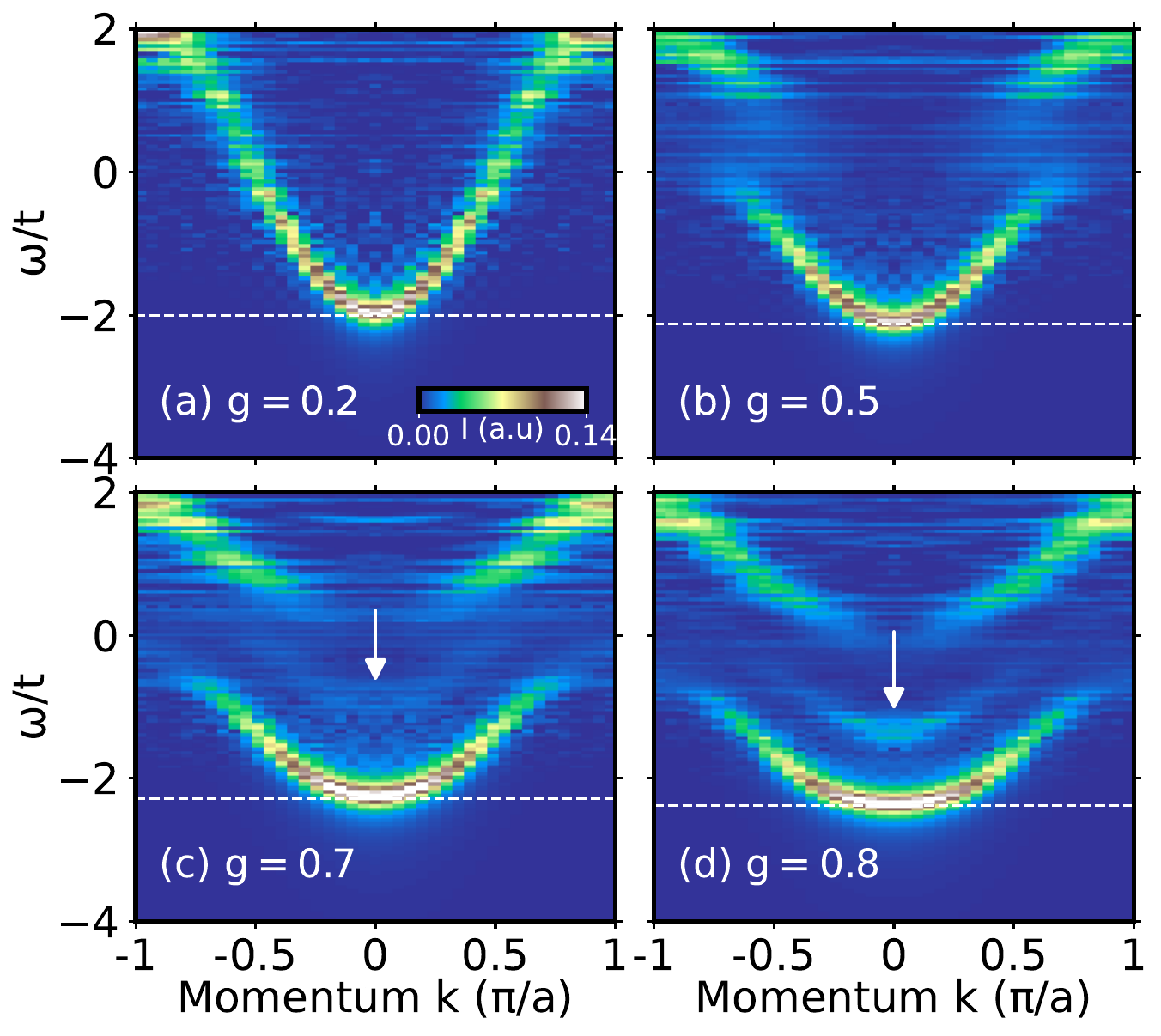}
    \caption{Electron addition spectral function $A^{+}(k,\omega)$ obtained with DMRG for adding a spin-$\downarrow$ electron to an $L=40$ sites \gls*{1D} chain occupied by one spin-$\uparrow$ electron. As indicated in each panel, results are shown for different values of $g$ while fixing $U=6t$, $V=0$, and $\Omega=2t$. The white dashed lines in each plot show the single polaron ground state energy, $E_{1,\mathrm{gs}}$. Arrows in panels c and d indicate the resonance that is the focus of this work.}
    \label{fig:Akw_vs_g}
\end{figure}

We solved Eq.~\eqref{eq:Hamiltonian_HSSH} using the \gls*{DMRG} method~\cite{DMRG_Steve_white_PRL}, as implemented in the DMRG++ code~\cite{DMRGpp}. In particular, we focus on the electron addition spectrum $A^+(k,\omega)$~\cite{damascelli2003angle,sobota2021angle}, which we compute directly on the real-frequency axis using the correction vector algorithm with Krylov space decomposition and a two-site \gls*{DMRG} update~\cite{Nocera2016spectralfunction}. The algorithm first computes the real space electron addition spectrum 
\begin{equation}
    A_{cj}^{+}(\omega) = -\frac{1}{\pi} \operatorname{Im} \bra{\psi_\mathrm{gs}} \hat{c}_{j,\sigma} 
    \frac{1}{\omega - \hat{H} + E_{\mathrm{gs}} + \text{i}\eta} 
    \hat{c}_{c,\sigma}^{\dagger} \ket{\psi_\mathrm{gs}},\label{eq:spectral_function}
\end{equation}
where $\ket{\psi_\mathrm{gs}}$ and $E_{\mathrm{gs}}$ are the ground state wave function and energy, and 
$c$ denotes the center site of the chain. The momentum-resolved electron addition spectrum $A^{+}(k,\omega)$ is then obtained by a Fourier transform. 

Throughout, we fixed the broadening coefficient to $\eta=0.1t$ and kept $m=500$ \gls*{DMRG} states while restricting the local phonon Hilbert space at most $N_p=7$ phonon modes per site to maintain a truncation error below $10^{-7}$.\\


\noindent{\bf Results} --- We first examine the electron addition spectra for a dilute \gls*{HSSH} chain with $U = 6t$ and $V = 0$. Figure~\ref{fig:Akw_vs_g} plots $A^+(k,\omega)$ for adding a spin-down electron to a chain already populated with one spin-up ground state polaron for different values of $g$. In this case, $E_{\mathrm{gs}}\equiv E_{1,\mathrm{gs}}$ in Eq.~\eqref{eq:spectral_function} refers to the single polaron ground state energy. This energy $E_{1,\mathrm{gs}}$ is indicated by the dashed white line and is equal to the Fermi energy $E_F$ in this case. Indeed, according to Eq.~\eqref{eq:spectral_function}, spectral weight is expected at energies $\omega = E_{2,\alpha}-E_{1,\mathrm{gs}}$ where $E_{2,\alpha}$ are the eigenenergies of $\hat{H}$ in the two-electron Hilbert space. Hence, in the absence of binding, the lowest-energy feature must appear at $\omega=E_{1,\mathrm{gs}}$ corresponding to $E_{2,\mathrm{gs}}= 2E_{1,\mathrm{gs}}$.

For weak coupling ($g=0.2$, Fig.~\ref{fig:Akw_vs_g}a), $A^+(k,\omega)$ indeed resembles the noninteracting cosine band, with the expected characteristic phonon kink appearing at $E_{1,\mathrm{gs}} + \Omega$~\cite{Engelsberg1963coupled, Bonca2019spectral}. Increasing \gls*{eph} coupling to $g=0.5$ (Fig.~\ref{fig:Akw_vs_g}b) produces a stronger renormalization and breaks the spectrum into a low-energy quasiparticle (polaron) discrete state and a higher-energy polaron $+$ one phonon continuum. Further increasing $g$ generates heavier polarons~\cite{hague2006effects,goodvin2006green}, and consequently reduces the bandwidth of the polaron band, as shown in Figs.~\ref{fig:Akw_vs_g}c and \ref{fig:Akw_vs_g}d. (For these coupling strengths, the large $U$ prevents the two polarons from binding together~\cite{bonca2000mobile,sous2017phonon, sous2018light}, hence there are no signatures of a bipolaron state.) For the stronger couplings, we observe the emergence of an unexpected anomalous spectral feature located within the polaron band, marked by the white arrows,  with larger \gls*{eph} couplings pushing its spectral weight closer to the bottom of the polaron band. As we will show next, this feature is a resonance arising from the residual $e$-$e$ attraction mediated by the \gls*{SSH} coupling~\cite{sous2018light}. 

\begin{figure}
    \centering
    \includegraphics[width=\columnwidth]{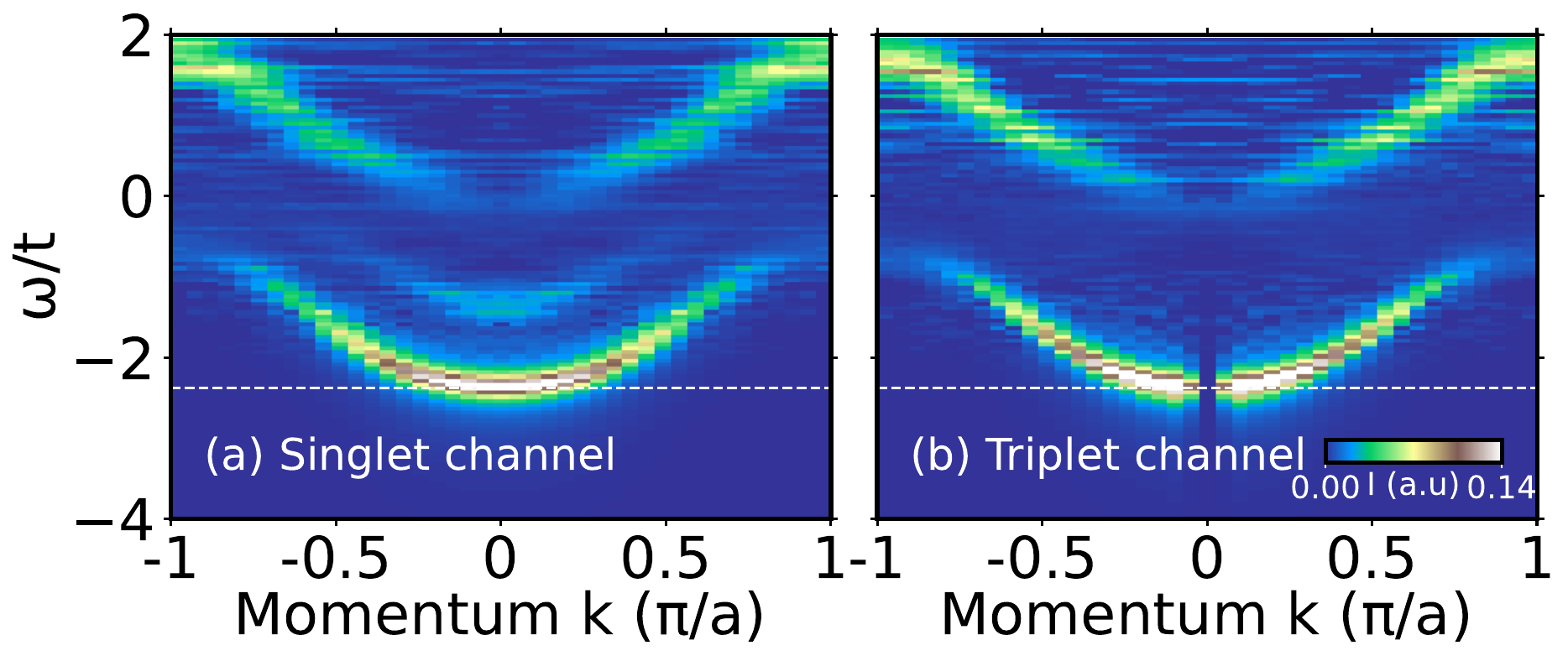}
    \caption{Comparison of electron addition spectral function $A^{+}(k,\omega)$ for adding either a) one spin-$\downarrow$ electron or b) one spin-$\uparrow$ electron to an $L=40$ site \gls*{1D} chain with a spin-$\uparrow$ electron already present. Results are obtained for fixed $U=6t$, $V=0$, $\Omega=2t$, and $g=0.8$. The white dashed lines in each plot show the single polaron ground state energy $E_{1,\mathrm{gs}}$.}
    \label{fig:singlet_vs_triplet}
\end{figure}

To further explore the nature of this anomalous spectral weight, Fig.~\ref{fig:singlet_vs_triplet} shows $A^{+}(k,\omega)$ in the singlet and triplet channels, obtained by adding a spin-down (singlet sector) or spin-up (triplet sector) electron to a system populated with a single spin-up polaron. Here, we set the parameters as in Fig.~\ref{fig:Akw_vs_g} and fix $g=0.8$. Notably, the anomalous spectral weight only appears in the spin-singlet channel, which can be rationalized by remembering that \gls*{SSH} coupling leads to phonon-mediated electron attraction (repulsion) in the spin-singlet (spin-triplet) channels \cite{sous2018light,sous2017phonon}. Thus, these results suggest that phonon-mediated pairing plays a significant role in the mechanisms responsible for this feature.

\begin{figure}[t]
    \centering
    \includegraphics[width=\columnwidth]{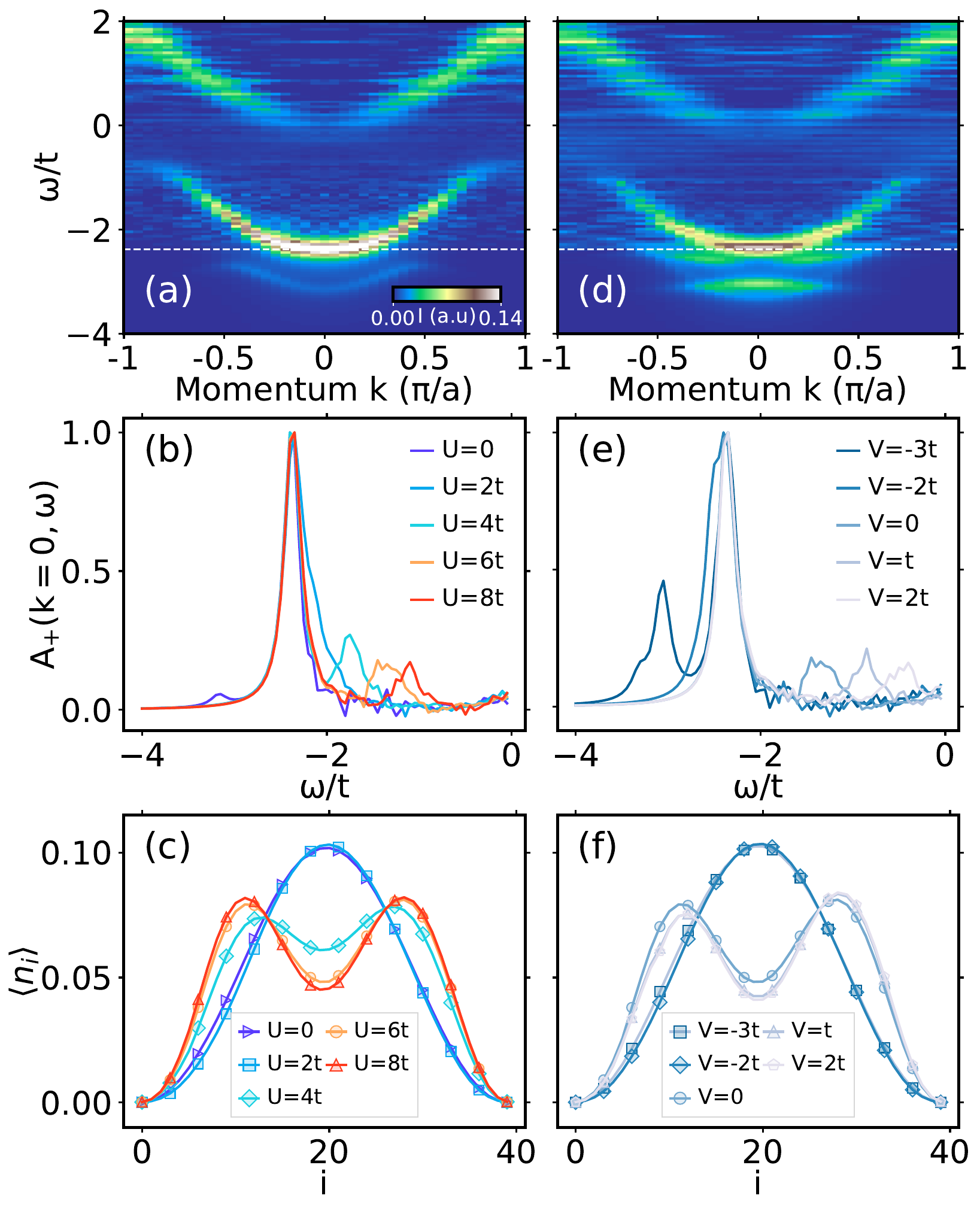}
    \caption{The addition spectrum $A^{+}(k,\omega)$ for adding a spin-$\downarrow$ electron to an $L=40$ site chain already occupied by a single spin-$\uparrow$ electron. We obtained all results for fixed $g = 0.8$ and $\Omega = 2t$. 
    Panels (a) and (d) show the momentum-resolved spectra for $(U, V) = (0,0)$ and $(6t,-3t)$, respectively. 
    Panel (b) plots the evolution of $A^{+}(k=0,\omega)$ with $U$ for fixed $V = 0$. Similarly, panel (e) plots $A^{+}(k=0,\omega)$ as a function of $V$ for fixed $U = 6t$. Panels (c) and (f) show average electron density $\langle n_i \rangle$ for the two-particle ground state obtained for the same parameters used in panels (b) and (e),  respectively.}
    \label{fig:Akw_vs_U_V}
\end{figure}

Further insights can be gained by examining how the addition spectrum evolves as a function of the Hubbard interaction parameters, as shown in Fig.~\ref{fig:Akw_vs_U_V} for fixed $g=0.8$. Fig.~\ref{fig:Akw_vs_U_V}a shows addition spectra for $U = V = 0$, which serves as a reference point, while Fig.~\ref{fig:Akw_vs_U_V}b plots the evolution of the $k=0$ spectra as a function of $U$ while fixing $V = 0$ and normalizing to the peak at $\omega \approx -2.5t$. As before, the dashed line in Fig.~\ref{fig:Akw_vs_U_V}a indicates the energy of the initial one-polaron state. 

For $U=0$, the \gls*{SSH} coupling is strong enough to bind the two electrons into a bound bipolaron, which manifests as the sharp spectral feature observed below the dashed line marking the $E_{1,\mathrm{gs}}$ in Fig.~\ref{fig:Akw_vs_U_V}a. Increasing $U$ reduces the bipolaron's binding energy and shifts its weight to higher energies~\cite{bonca2000mobile}. For $U = 2t$, we find that the bipolaron peak has already merged with the 
bottom of the polaron band, as indicated by the asymmetric line shape in Fig.~\ref{fig:Akw_vs_U_V}b. Thus, for $U\ge 2t$, the bipolaron becomes unstable to dissociate into two free polarons, because its energy is larger than $2E_{1,\mathrm{gs}}$. However, a higher-energy finite-lifetime resonance can persist in the spectrum due to the residual strong attraction between the carriers. We interpret the sharp spectral feature above the bottom of the polaron band as such a resonance, which is shifted to higher energies with increasing $U$. The change in the character of the two-electron ground state is also reflected in the average densities calculated for the two-electron ground state shown in Fig.~\ref{fig:Akw_vs_U_V}c, where we observe a change from a single peak (bound bipolaron) to a bimodal distribution (two unbound polarons) for $U > 2t$~\cite{nocera2021bipolaron}. 

Figures~\ref{fig:Akw_vs_U_V}(d)-(f) illustrate the response of the two-electron state as a function of nearest-neighbor interaction $V$. Fig.~\ref{fig:Akw_vs_U_V}d shows $A^{+}(k,\omega)$ for fixed $U=6t$, $g=0.8$, and $V=-3t$. Similar to Fig.~\ref{fig:Akw_vs_U_V}a, the ground state is a bound bipolaron, which is again reflected by the formation of a sharp spectral feature below the dashed line. Sweeping $V$ from attractive to repulsive values reduces the bipolaron binding energy, similar to the effect of increasing $U$. For $V>0$, the bipolaron is unstable to dissociation, but the remnant phonon-mediated attraction produces a resonance, similar to Fig.~\ref{fig:Akw_vs_U_V}b. The change from the bipolaron to the two unbound polaron ground state is also clear from the behavior of the average two-electron densities as a function of $V$, as shown in Fig.~\ref{fig:Akw_vs_U_V}f.

To confirm our interpretation of the numerical results, we conducted a perturbative analysis of our model in the anti-adiabatic limit ($\Omega \gg t, g)$, as described in the \gls*{SM}~\cite{Supplement}. Specifically, we use M. Takahashi's method~\cite{Takahashi77} to project out the high-energy Hilbert subspaces with one or more phonons to obtain an effective two-particle Hamiltonian $H_{\rm eff}$ for the \gls*{HSSH} model. 
In the singlet sector, $H_{\rm eff}$ has terms 
describing the renormalized single quasiparticle (polaron) dispersion plus terms describing their interactions; the latter includes both the bare on-site repulsion $U$ and the phonon-mediated attraction. The latter consists of nearest-neighbor pair hopping of on-site and nearest-neighbor singlets~\cite{Supplement,sous2018light}. In momentum space, these phonon-mediated interactions describe the scattering of a singlet  $|k,q\rangle ={1\over \sqrt{2}}[c^\dagger_{{k\over 2}+q, \uparrow}c^\dagger_{{k\over 2}-q, \downarrow}-c^\dagger_{{k\over 2}-q, \uparrow} c^\dagger_{{k\over 2}+q, \downarrow}]|0\rangle$  into any other singlet $|k,q'\rangle$, in terms of on-site $u_0(k)$, nearest-neighbor $u_1(k)\cos(qa)\cos(q'a)$, and next nearest-neighbor $u_2(k)[\cos(2qa)+\cos(2q^\prime a)]$ effective interactions which depend not only on the transferred momentum $q'-q$ but also on the total momentum $k$ of the pair, see the \gls*{SM} for details~\cite{Supplement}.

We then mimic the spectral weight of Eq.~\eqref{eq:spectral_function} (up to the shift by $E_{1,\mathrm{gs}}$)  by calculating the two-particle propagator $G(k,z,\tfrac{k}{2},\tfrac{k}2)=\langle k, {k\over 2}|[z-H_{\rm eff}]^{-1}|k, {k\over 2}\rangle$ corresponding to a singlet between a particle with momentum $k$ and one with zero momentum. As shown in the \gls*{SM}~\cite{Supplement}, this spectral weight behaves qualitatively similarly to $A^+(k,\omega)$ of the DMRG simulations. In particular, it exhibits a similar resonance at appropriate $g,U$ values. The analytical approximation reveals that in the anti-adiabatic limit, the bipolaron is stable, like in Fig.~\ref{fig:Akw_vs_U_V}(a), when the total on-site effective energy $U+ u_0(k) = U-\frac{8g^2}{\Omega} \cos(ka)<0$, {\em i.e.} when the Hubbard repulsion is weaker than the on-site phonon-mediated attraction (the latter depends on the momentum $k$ of the pair because it arises from a pair-hopping term). However, for $U>2t$ and $g=0.8$, $u_0(k)$ becomes repulsive and there are no stable bipolarons [see Fig.~\ref{fig:Akw_vs_U_V}b]. Nevertheless, the nearest- and next-nearest neighbor $u_1(k)$ and $u_2(k)$ are still attractive near $k=0$. While they are not sufficiently strong to bind a stable bipolaron, they do mediate the appearance of the finite lifetime resonance, seen in Fig.~\ref{fig:Akw_vs_U_V}b and also shown in the \gls*{SM}~\cite{Supplement}.

\begin{figure}[t]
    \centering
    \includegraphics[width=\columnwidth]{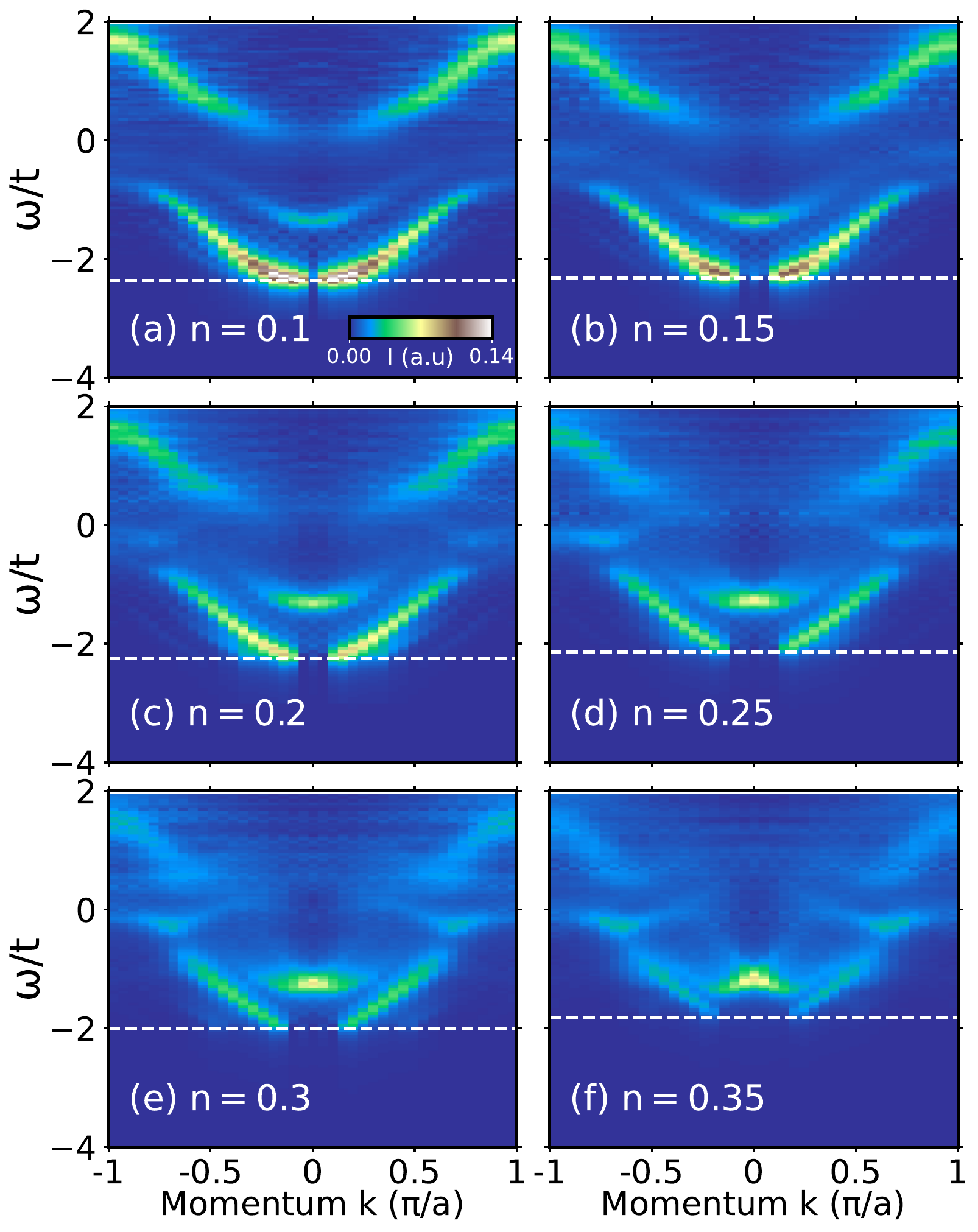}
    \caption{Electron addition spectral function $A^+(k,\omega)$ calculated with \gls*{DMRG} for an $L=40$ site chain and fixed $U=6t$, $V=0$, $g=0.8$, and $\Omega=2t$. The average electron density $\langle n \rangle$ is varied between $0.1$ and $0.35$, as indicated in each panel. Here, we find that the resonance feature persists well beyond the dilute limit.}
    \label{fig:Akw_vs_density}
\end{figure}

The anomalous spectral feature discussed above provides a clear spectral signature of the existence of the non-local attractive interaction mediated by the \gls*{SSH} interaction, first predicted in Ref.~\cite{sous2018light}, even when this is not strong enough to bind a bipolaron. Given this, it is natural to ask whether this feature persists beyond the extremely dilute limit and could be observed for carrier concentrations relevant to most quantum materials. Fig.~\ref{fig:Akw_vs_density} answers this question by plotting electron addition spectra for $U = 6t$, $V = 0$, and $g = 0.8$ for carrier concentrations ranging from $\langle n \rangle \in [0.1, 0.35]$. The resonance is visible in all cases, with a spectral weight increasing monotonically with $\langle n \rangle$. This increase is expected because the added electron can form a resonance with any of the electrons with opposite spin already present in the system. For $\langle n \rangle\ge 0.2$, this feature develops additional structure and a W-like shape that places the minima away from the $\Gamma$-point. For the largest carrier concentrations [see Fig.~\ref{fig:Akw_vs_density}f], the feature becomes very reminiscent of the spectral function of the lightly doped \gls*{1D} Hubbard model, where spin-charge separation occurs with distinct holon and spinon quasiparticle excitations. \\

\noindent{\bf Summary \& Discussion} --- We have calculated the electron addition spectrum for an extended-\gls*{HSSH} model for carrier concentrations $\langle n \rangle \le 0.35$ and identified an anomalous spectral feature that lays within the two polaron continuum. Through careful analysis, we have demonstrated that this feature is a resonance state where the injected electron forms a metastable singlet bound state for some time, before decaying. (This interpretation entirely agrees with an analytical treatment of the problem in the anti-adiabatic limit~\cite{Supplement}.) This feature's precise shape and location depend on the on-site and next-nearest neighbor Hubbard interactions and the strength of the \gls*{eph} coupling and encodes information about the residual phonon-mediated attraction in the \gls*{HSSH} model. Calculations for doped-\gls*{HSSH} chains with $\langle n \rangle = 0.75$ and $\Omega = 2t$ did not observe any resonance feature; instead, the reported $A(k,\omega)$ exhibited the canonical features of spin-charge separation~\cite{Banerjee2023groundstate}. Combined with our results, this observation suggests that the effective attraction between the polarons is ultimately suppressed at higher carrier concentrations. 

As highlighted earlier, previous theoretical work has demonstrated that an \gls*{SSH} coupling mediates a strong attractive interaction between carriers in the dilute limit, forming light but stable bipolarons~\cite{sous2018light}. Our \gls*{DMRG} results indicate that this interaction persists beyond the dilute limit and provides a direct spectroscopic signature of this interaction accessible to probes accessing the \textit{unoccupied} electronic states. Scanning tunneling microscopy and inverse photoemission experiments are directly suited for this purpose; however, it would also be interesting to determine if \gls*{RIXS} can access these states. In this context, we note that doped \gls*{1D} cuprate spin changes have recently been synthesized and characterized~\cite{Chen2021anomalously} with estimated doping levels as low as $\rho = 1-\langle n \rangle =0.59$ holes/Cu. Higher doping levels need to be achieved to test our predictions. Analogous calculations in higher dimensions would also help identify suitable materials. \\

\noindent{\bf Acknowledgments}: This work was supported by the National Science Foundation under Grant No. DMR-2401388. A.~N. acknowledges the support of the Canada First Research Excellence Fund. This research used resources from the Oak Ridge Leadership Computing Facility, which is a DOE Office of Science User Facility supported under Contract No. DE-AC05-00OR22725. \\

\noindent{\bf Data Availability}: The data supporting this study will be deposited in a public online repository once the paper is accepted. Until that time, data will be made available upon request. 
\bibliography{references.bib}

\begin{thebibliography}{54}%
\makeatletter
\providecommand \@ifxundefined [1]{%
 \@ifx{#1\undefined}
}%
\providecommand \@ifnum [1]{%
 \ifnum #1\expandafter \@firstoftwo
 \else \expandafter \@secondoftwo
 \fi
}%
\providecommand \@ifx [1]{%
 \ifx #1\expandafter \@firstoftwo
 \else \expandafter \@secondoftwo
 \fi
}%
\providecommand \natexlab [1]{#1}%
\providecommand \enquote  [1]{``#1''}%
\providecommand \bibnamefont  [1]{#1}%
\providecommand \bibfnamefont [1]{#1}%
\providecommand \citenamefont [1]{#1}%
\providecommand \href@noop [0]{\@secondoftwo}%
\providecommand \href [0]{\begingroup \@sanitize@url \@href}%
\providecommand \@href[1]{\@@startlink{#1}\@@href}%
\providecommand \@@href[1]{\endgroup#1\@@endlink}%
\providecommand \@sanitize@url [0]{\catcode `\\12\catcode `\$12\catcode `\&12\catcode `\#12\catcode `\^12\catcode `\_12\catcode `\%12\relax}%
\providecommand \@@startlink[1]{}%
\providecommand \@@endlink[0]{}%
\providecommand \url  [0]{\begingroup\@sanitize@url \@url }%
\providecommand \@url [1]{\endgroup\@href {#1}{\urlprefix }}%
\providecommand \urlprefix  [0]{URL }%
\providecommand \Eprint [0]{\href }%
\providecommand \doibase [0]{https://doi.org/}%
\providecommand \selectlanguage [0]{\@gobble}%
\providecommand \bibinfo  [0]{\@secondoftwo}%
\providecommand \bibfield  [0]{\@secondoftwo}%
\providecommand \translation [1]{[#1]}%
\providecommand \BibitemOpen [0]{}%
\providecommand \bibitemStop [0]{}%
\providecommand \bibitemNoStop [0]{.\EOS\space}%
\providecommand \EOS [0]{\spacefactor3000\relax}%
\providecommand \BibitemShut  [1]{\csname bibitem#1\endcsname}%
\let\auto@bib@innerbib\@empty
\bibitem [{\citenamefont {Ashcroft}(1968)}]{Achroft1968metallic}%
  \BibitemOpen
  \bibfield  {author} {\bibinfo {author} {\bibfnamefont {N.~W.}\ \bibnamefont {Ashcroft}},\ }\bibfield  {title} {\bibinfo {title} {Metallic hydrogen: A high-temperature superconductor?},\ }\href {https://doi.org/10.1103/PhysRevLett.21.1748} {\bibfield  {journal} {\bibinfo  {journal} {Phys. Rev. Lett.}\ }\textbf {\bibinfo {volume} {21}},\ \bibinfo {pages} {1748} (\bibinfo {year} {1968})}\BibitemShut {NoStop}%
\bibitem [{\citenamefont {Cohen}\ and\ \citenamefont {Anderson}(1972)}]{Cohen1972comments}%
  \BibitemOpen
  \bibfield  {author} {\bibinfo {author} {\bibfnamefont {M.~L.}\ \bibnamefont {Cohen}}\ and\ \bibinfo {author} {\bibfnamefont {P.~W.}\ \bibnamefont {Anderson}},\ }\bibfield  {title} {\bibinfo {title} {Comments on the maximum superconducting transition temperature},\ }\href {https://doi.org/10.1063/1.2946185} {\bibfield  {journal} {\bibinfo  {journal} {AIP Conference Proceedings}\ }\textbf {\bibinfo {volume} {4}},\ \bibinfo {pages} {17} (\bibinfo {year} {1972})}\BibitemShut {NoStop}%
\bibitem [{\citenamefont {Allen}\ and\ \citenamefont {Mitrovi{\'c}}(1983)}]{Allen1983theory}%
  \BibitemOpen
  \bibfield  {author} {\bibinfo {author} {\bibfnamefont {P.~B.}\ \bibnamefont {Allen}}\ and\ \bibinfo {author} {\bibfnamefont {B.}~\bibnamefont {Mitrovi{\'c}}},\ }\bibfield  {title} {\bibinfo {title} {Theory of superconducting {T$_\mathrm{c}$}}\ }(\bibinfo  {publisher} {Academic Press},\ \bibinfo {year} {1983})\ pp.\ \bibinfo {pages} {1--92}\BibitemShut {NoStop}%
\bibitem [{\citenamefont {Chakraverty}\ \emph {et~al.}(1998)\citenamefont {Chakraverty}, \citenamefont {Ranninger},\ and\ \citenamefont {Feinberg}}]{chakraverty1998experimental}%
  \BibitemOpen
  \bibfield  {author} {\bibinfo {author} {\bibfnamefont {B.~K.}\ \bibnamefont {Chakraverty}}, \bibinfo {author} {\bibfnamefont {J.}~\bibnamefont {Ranninger}},\ and\ \bibinfo {author} {\bibfnamefont {D.}~\bibnamefont {Feinberg}},\ }\bibfield  {title} {\bibinfo {title} {Experimental and theoretical constraints of bipolaronic superconductivity in high {${T}_{c}$} materials: An impossibility},\ }\href {https://doi.org/10.1103/PhysRevLett.81.433} {\bibfield  {journal} {\bibinfo  {journal} {Phys. Rev. Lett.}\ }\textbf {\bibinfo {volume} {81}},\ \bibinfo {pages} {433} (\bibinfo {year} {1998})}\BibitemShut {NoStop}%
\bibitem [{\citenamefont {Kuli\ifmmode~\acute{c}\else \'{c}\fi{}}\ and\ \citenamefont {Zeyher}(1994)}]{Kulic1994influence}%
  \BibitemOpen
  \bibfield  {author} {\bibinfo {author} {\bibfnamefont {M.~L.}\ \bibnamefont {Kuli\ifmmode~\acute{c}\else \'{c}\fi{}}}\ and\ \bibinfo {author} {\bibfnamefont {R.}~\bibnamefont {Zeyher}},\ }\bibfield  {title} {\bibinfo {title} {Influence of strong electron correlations on the electron-phonon coupling in high-{${\mathit{T}}_{\mathit{c}}$} oxides},\ }\href {https://doi.org/10.1103/PhysRevB.49.4395} {\bibfield  {journal} {\bibinfo  {journal} {Phys. Rev. B}\ }\textbf {\bibinfo {volume} {49}},\ \bibinfo {pages} {4395} (\bibinfo {year} {1994})}\BibitemShut {NoStop}%
\bibitem [{\citenamefont {Kuli{\'c}}(2000)}]{Kulic2000interplay}%
  \BibitemOpen
  \bibfield  {author} {\bibinfo {author} {\bibfnamefont {M.~L.}\ \bibnamefont {Kuli{\'c}}},\ }\bibfield  {title} {\bibinfo {title} {Interplay of electron--phonon interaction and strong correlations: the possible way to high-temperature superconductivity},\ }\href {https://doi.org/https://doi.org/10.1016/S0370-1573(00)00008-9} {\bibfield  {journal} {\bibinfo  {journal} {Physics Reports}\ }\textbf {\bibinfo {volume} {338}},\ \bibinfo {pages} {1} (\bibinfo {year} {2000})}\BibitemShut {NoStop}%
\bibitem [{\citenamefont {Moussa}\ and\ \citenamefont {Cohen}(2006)}]{Moussa2006twobounds}%
  \BibitemOpen
  \bibfield  {author} {\bibinfo {author} {\bibfnamefont {J.~E.}\ \bibnamefont {Moussa}}\ and\ \bibinfo {author} {\bibfnamefont {M.~L.}\ \bibnamefont {Cohen}},\ }\bibfield  {title} {\bibinfo {title} {Two bounds on the maximum phonon-mediated superconducting transition temperature},\ }\href {https://doi.org/10.1103/PhysRevB.74.094520} {\bibfield  {journal} {\bibinfo  {journal} {Phys. Rev. B}\ }\textbf {\bibinfo {volume} {74}},\ \bibinfo {pages} {094520} (\bibinfo {year} {2006})}\BibitemShut {NoStop}%
\bibitem [{\citenamefont {Gor'kov}\ and\ \citenamefont {Kresin}(2018)}]{Gorkov2018colloquium}%
  \BibitemOpen
  \bibfield  {author} {\bibinfo {author} {\bibfnamefont {L.~P.}\ \bibnamefont {Gor'kov}}\ and\ \bibinfo {author} {\bibfnamefont {V.~Z.}\ \bibnamefont {Kresin}},\ }\bibfield  {title} {\bibinfo {title} {Colloquium: High pressure and road to room temperature superconductivity},\ }\href {https://doi.org/10.1103/RevModPhys.90.011001} {\bibfield  {journal} {\bibinfo  {journal} {Rev. Mod. Phys.}\ }\textbf {\bibinfo {volume} {90}},\ \bibinfo {pages} {011001} (\bibinfo {year} {2018})}\BibitemShut {NoStop}%
\bibitem [{\citenamefont {Esterlis}\ \emph {et~al.}(2018{\natexlab{a}})\citenamefont {Esterlis}, \citenamefont {Kivelson},\ and\ \citenamefont {Scalapino}}]{Esterlis2018bound}%
  \BibitemOpen
  \bibfield  {author} {\bibinfo {author} {\bibfnamefont {I.}~\bibnamefont {Esterlis}}, \bibinfo {author} {\bibfnamefont {S.~A.}\ \bibnamefont {Kivelson}},\ and\ \bibinfo {author} {\bibfnamefont {D.~J.}\ \bibnamefont {Scalapino}},\ }\bibfield  {title} {\bibinfo {title} {A bound on the superconducting transition temperature},\ }\href {https://doi.org/10.1038/s41535-018-0133-0} {\bibfield  {journal} {\bibinfo  {journal} {npj Quantum Materials}\ }\textbf {\bibinfo {volume} {3}},\ \bibinfo {pages} {59} (\bibinfo {year} {2018}{\natexlab{a}})}\BibitemShut {NoStop}%
\bibitem [{\citenamefont {Esterlis}\ \emph {et~al.}(2018{\natexlab{b}})\citenamefont {Esterlis}, \citenamefont {Nosarzewski}, \citenamefont {Huang}, \citenamefont {Moritz}, \citenamefont {Devereaux}, \citenamefont {Scalapino},\ and\ \citenamefont {Kivelson}}]{Esterlis2018breakdown}%
  \BibitemOpen
  \bibfield  {author} {\bibinfo {author} {\bibfnamefont {I.}~\bibnamefont {Esterlis}}, \bibinfo {author} {\bibfnamefont {B.}~\bibnamefont {Nosarzewski}}, \bibinfo {author} {\bibfnamefont {E.~W.}\ \bibnamefont {Huang}}, \bibinfo {author} {\bibfnamefont {B.}~\bibnamefont {Moritz}}, \bibinfo {author} {\bibfnamefont {T.~P.}\ \bibnamefont {Devereaux}}, \bibinfo {author} {\bibfnamefont {D.~J.}\ \bibnamefont {Scalapino}},\ and\ \bibinfo {author} {\bibfnamefont {S.~A.}\ \bibnamefont {Kivelson}},\ }\bibfield  {title} {\bibinfo {title} {Breakdown of the {M}igdal-{E}liashberg theory: A determinant quantum {M}onte {C}arlo study},\ }\href {https://doi.org/10.1103/PhysRevB.97.140501} {\bibfield  {journal} {\bibinfo  {journal} {Phys. Rev. B}\ }\textbf {\bibinfo {volume} {97}},\ \bibinfo {pages} {140501} (\bibinfo {year} {2018}{\natexlab{b}})}\BibitemShut {NoStop}%
\bibitem [{\citenamefont {Hazra}\ \emph {et~al.}(2019)\citenamefont {Hazra}, \citenamefont {Verma},\ and\ \citenamefont {Randeria}}]{Hazra2019bounds}%
  \BibitemOpen
  \bibfield  {author} {\bibinfo {author} {\bibfnamefont {T.}~\bibnamefont {Hazra}}, \bibinfo {author} {\bibfnamefont {N.}~\bibnamefont {Verma}},\ and\ \bibinfo {author} {\bibfnamefont {M.}~\bibnamefont {Randeria}},\ }\bibfield  {title} {\bibinfo {title} {Bounds on the superconducting transition temperature: Applications to twisted bilayer graphene and cold atoms},\ }\href {https://doi.org/10.1103/PhysRevX.9.031049} {\bibfield  {journal} {\bibinfo  {journal} {Phys. Rev. X}\ }\textbf {\bibinfo {volume} {9}},\ \bibinfo {pages} {031049} (\bibinfo {year} {2019})}\BibitemShut {NoStop}%
\bibitem [{\citenamefont {Esterlis}\ \emph {et~al.}(2019)\citenamefont {Esterlis}, \citenamefont {Kivelson},\ and\ \citenamefont {Scalapino}}]{Esterlis2019pseudogap}%
  \BibitemOpen
  \bibfield  {author} {\bibinfo {author} {\bibfnamefont {I.}~\bibnamefont {Esterlis}}, \bibinfo {author} {\bibfnamefont {S.~A.}\ \bibnamefont {Kivelson}},\ and\ \bibinfo {author} {\bibfnamefont {D.~J.}\ \bibnamefont {Scalapino}},\ }\bibfield  {title} {\bibinfo {title} {Pseudogap crossover in the electron-phonon system},\ }\href {https://doi.org/10.1103/PhysRevB.99.174516} {\bibfield  {journal} {\bibinfo  {journal} {Phys. Rev. B}\ }\textbf {\bibinfo {volume} {99}},\ \bibinfo {pages} {174516} (\bibinfo {year} {2019})}\BibitemShut {NoStop}%
\bibitem [{\citenamefont {Hofmann}\ \emph {et~al.}(2022)\citenamefont {Hofmann}, \citenamefont {Chowdhury}, \citenamefont {Kivelson},\ and\ \citenamefont {Berg}}]{Hofmann2022heuristic}%
  \BibitemOpen
  \bibfield  {author} {\bibinfo {author} {\bibfnamefont {J.~S.}\ \bibnamefont {Hofmann}}, \bibinfo {author} {\bibfnamefont {D.}~\bibnamefont {Chowdhury}}, \bibinfo {author} {\bibfnamefont {S.~A.}\ \bibnamefont {Kivelson}},\ and\ \bibinfo {author} {\bibfnamefont {E.}~\bibnamefont {Berg}},\ }\bibfield  {title} {\bibinfo {title} {Heuristic bounds on superconductivity and how to exceed them},\ }\href {https://doi.org/10.1038/s41535-022-00491-1} {\bibfield  {journal} {\bibinfo  {journal} {npj Quantum Materials}\ }\textbf {\bibinfo {volume} {7}},\ \bibinfo {pages} {83} (\bibinfo {year} {2022})}\BibitemShut {NoStop}%
\bibitem [{\citenamefont {Pickett}(2023)}]{Pickett2023colloquium}%
  \BibitemOpen
  \bibfield  {author} {\bibinfo {author} {\bibfnamefont {W.~E.}\ \bibnamefont {Pickett}},\ }\bibfield  {title} {\bibinfo {title} {Colloquium: Room temperature superconductivity: The roles of theory and materials design},\ }\href {https://doi.org/10.1103/RevModPhys.95.021001} {\bibfield  {journal} {\bibinfo  {journal} {Rev. Mod. Phys.}\ }\textbf {\bibinfo {volume} {95}},\ \bibinfo {pages} {021001} (\bibinfo {year} {2023})}\BibitemShut {NoStop}%
\bibitem [{\citenamefont {Nosarzewski}\ \emph {et~al.}(2021)\citenamefont {Nosarzewski}, \citenamefont {Huang}, \citenamefont {Dee}, \citenamefont {Esterlis}, \citenamefont {Moritz}, \citenamefont {Kivelson}, \citenamefont {Johnston},\ and\ \citenamefont {Devereaux}}]{Nosarzewski2021superconductivity}%
  \BibitemOpen
  \bibfield  {author} {\bibinfo {author} {\bibfnamefont {B.}~\bibnamefont {Nosarzewski}}, \bibinfo {author} {\bibfnamefont {E.~W.}\ \bibnamefont {Huang}}, \bibinfo {author} {\bibfnamefont {P.~M.}\ \bibnamefont {Dee}}, \bibinfo {author} {\bibfnamefont {I.}~\bibnamefont {Esterlis}}, \bibinfo {author} {\bibfnamefont {B.}~\bibnamefont {Moritz}}, \bibinfo {author} {\bibfnamefont {S.~A.}\ \bibnamefont {Kivelson}}, \bibinfo {author} {\bibfnamefont {S.}~\bibnamefont {Johnston}},\ and\ \bibinfo {author} {\bibfnamefont {T.~P.}\ \bibnamefont {Devereaux}},\ }\bibfield  {title} {\bibinfo {title} {Superconductivity, charge density waves, and bipolarons in the {H}olstein model},\ }\href {https://doi.org/10.1103/PhysRevB.103.235156} {\bibfield  {journal} {\bibinfo  {journal} {Phys. Rev. B}\ }\textbf {\bibinfo {volume} {103}},\ \bibinfo {pages} {235156} (\bibinfo {year} {2021})}\BibitemShut {NoStop}%
\bibitem [{\citenamefont {Bradley}\ \emph {et~al.}(2021)\citenamefont {Bradley}, \citenamefont {Batrouni},\ and\ \citenamefont {Scalettar}}]{Bradley2021superconductivity}%
  \BibitemOpen
  \bibfield  {author} {\bibinfo {author} {\bibfnamefont {O.}~\bibnamefont {Bradley}}, \bibinfo {author} {\bibfnamefont {G.~G.}\ \bibnamefont {Batrouni}},\ and\ \bibinfo {author} {\bibfnamefont {R.~T.}\ \bibnamefont {Scalettar}},\ }\bibfield  {title} {\bibinfo {title} {Superconductivity and charge density wave order in the two-dimensional {H}olstein model},\ }\href {https://doi.org/10.1103/PhysRevB.103.235104} {\bibfield  {journal} {\bibinfo  {journal} {Phys. Rev. B}\ }\textbf {\bibinfo {volume} {103}},\ \bibinfo {pages} {235104} (\bibinfo {year} {2021})}\BibitemShut {NoStop}%
\bibitem [{\citenamefont {Zhang}\ \emph {et~al.}(2023)\citenamefont {Zhang}, \citenamefont {Sous}, \citenamefont {Reichman}, \citenamefont {Berciu}, \citenamefont {Millis}, \citenamefont {Prokof'ev},\ and\ \citenamefont {Svistunov}}]{zhang2023bipolaronic}%
  \BibitemOpen
  \bibfield  {author} {\bibinfo {author} {\bibfnamefont {C.}~\bibnamefont {Zhang}}, \bibinfo {author} {\bibfnamefont {J.}~\bibnamefont {Sous}}, \bibinfo {author} {\bibfnamefont {D.~R.}\ \bibnamefont {Reichman}}, \bibinfo {author} {\bibfnamefont {M.}~\bibnamefont {Berciu}}, \bibinfo {author} {\bibfnamefont {A.~J.}\ \bibnamefont {Millis}}, \bibinfo {author} {\bibfnamefont {N.~V.}\ \bibnamefont {Prokof'ev}},\ and\ \bibinfo {author} {\bibfnamefont {B.~V.}\ \bibnamefont {Svistunov}},\ }\bibfield  {title} {\bibinfo {title} {Bipolaronic high-temperature superconductivity},\ }\href {https://doi.org/10.1103/PhysRevX.13.011010} {\bibfield  {journal} {\bibinfo  {journal} {Phys. Rev. X}\ }\textbf {\bibinfo {volume} {13}},\ \bibinfo {pages} {011010} (\bibinfo {year} {2023})}\BibitemShut {NoStop}%
\bibitem [{\citenamefont {Semenok}\ \emph {et~al.}(2024)\citenamefont {Semenok}, \citenamefont {Altshuler},\ and\ \citenamefont {Yuzbashyan}}]{semenok2024fundamental}%
  \BibitemOpen
  \bibfield  {author} {\bibinfo {author} {\bibfnamefont {D.~V.}\ \bibnamefont {Semenok}}, \bibinfo {author} {\bibfnamefont {B.~L.}\ \bibnamefont {Altshuler}},\ and\ \bibinfo {author} {\bibfnamefont {E.~A.}\ \bibnamefont {Yuzbashyan}},\ }\bibfield  {title} {\bibinfo {title} {Fundamental limits on the electron-phonon coupling and superconducting {$T_c$}},\ }\href {https://arxiv.org/abs/2407.12922} {\bibfield  {journal} {\bibinfo  {journal} {arXiv:2407.12922}\ } (\bibinfo {year} {2024})}\BibitemShut {NoStop}%
\bibitem [{\citenamefont {Trachenko}\ \emph {et~al.}(2025)\citenamefont {Trachenko}, \citenamefont {Monserrat}, \citenamefont {Hutcheon},\ and\ \citenamefont {Pickard}}]{trachenko2025upper}%
  \BibitemOpen
  \bibfield  {author} {\bibinfo {author} {\bibfnamefont {K.}~\bibnamefont {Trachenko}}, \bibinfo {author} {\bibfnamefont {B.}~\bibnamefont {Monserrat}}, \bibinfo {author} {\bibfnamefont {M.}~\bibnamefont {Hutcheon}},\ and\ \bibinfo {author} {\bibfnamefont {C.~J.}\ \bibnamefont {Pickard}},\ }\bibfield  {title} {\bibinfo {title} {Upper bounds on the highest phonon frequency and superconducting temperature from fundamental physical constants},\ }\href {https://arxiv.org/abs/2406.08129} {\bibfield  {journal} {\bibinfo  {journal} {arXiv:2406.08129}\ } (\bibinfo {year} {2025})}\BibitemShut {NoStop}%
\bibitem [{\citenamefont {Eliashberg}(1960)}]{Eliashberg}%
  \BibitemOpen
  \bibfield  {author} {\bibinfo {author} {\bibfnamefont {G.~M.}\ \bibnamefont {Eliashberg}},\ }\bibfield  {title} {\bibinfo {title} {Interactions between electrons and lattice vibrations in a superconductor},\ }\href {http://jetp.ras.ru/cgi-bin/e/index/e/11/3/p696?a=list} {\bibfield  {journal} {\bibinfo  {journal} {Zh. Eksp. Teor. Fiz.}\ }\textbf {\bibinfo {volume} {38}},\ \bibinfo {pages} {966} (\bibinfo {year} {1960})}\BibitemShut {NoStop}%
\bibitem [{\citenamefont {Allen}\ and\ \citenamefont {Dynes}(1975)}]{Allen1975transition}%
  \BibitemOpen
  \bibfield  {author} {\bibinfo {author} {\bibfnamefont {P.~B.}\ \bibnamefont {Allen}}\ and\ \bibinfo {author} {\bibfnamefont {R.~C.}\ \bibnamefont {Dynes}},\ }\bibfield  {title} {\bibinfo {title} {Transition temperature of strong-coupled superconductors reanalyzed},\ }\href {https://doi.org/10.1103/PhysRevB.12.905} {\bibfield  {journal} {\bibinfo  {journal} {Phys. Rev. B}\ }\textbf {\bibinfo {volume} {12}},\ \bibinfo {pages} {905} (\bibinfo {year} {1975})}\BibitemShut {NoStop}%
\bibitem [{\citenamefont {Issa}\ \emph {et~al.}(2025)\citenamefont {Issa}, \citenamefont {Bradley}, \citenamefont {Khatami},\ and\ \citenamefont {Scalettar}}]{issa2025learningconfusionphasediagram}%
  \BibitemOpen
  \bibfield  {author} {\bibinfo {author} {\bibfnamefont {G.}~\bibnamefont {Issa}}, \bibinfo {author} {\bibfnamefont {O.}~\bibnamefont {Bradley}}, \bibinfo {author} {\bibfnamefont {E.}~\bibnamefont {Khatami}},\ and\ \bibinfo {author} {\bibfnamefont {R.}~\bibnamefont {Scalettar}},\ }\bibfield  {title} {\bibinfo {title} {Learning by confusion: The phase diagram of the {H}olstein model},\ }\href {https://arxiv.org/abs/2501.04681} {\bibfield  {journal} {\bibinfo  {journal} {arXiv:2501.04681}\ } (\bibinfo {year} {2025})}\BibitemShut {NoStop}%
\bibitem [{\citenamefont {Bari\ifmmode \check{s}\else \v{s}\fi{}i\ifmmode~\acute{c}\else \'{c}\fi{}}\ \emph {et~al.}(1970)\citenamefont {Bari\ifmmode \check{s}\else \v{s}\fi{}i\ifmmode~\acute{c}\else \'{c}\fi{}}, \citenamefont {Labb\'e},\ and\ \citenamefont {Friedel}}]{Barisic1970tightbinding}%
  \BibitemOpen
  \bibfield  {author} {\bibinfo {author} {\bibfnamefont {S.}~\bibnamefont {Bari\ifmmode \check{s}\else \v{s}\fi{}i\ifmmode~\acute{c}\else \'{c}\fi{}}}, \bibinfo {author} {\bibfnamefont {J.}~\bibnamefont {Labb\'e}},\ and\ \bibinfo {author} {\bibfnamefont {J.}~\bibnamefont {Friedel}},\ }\bibfield  {title} {\bibinfo {title} {Tight binding and transition-metal superconductivity},\ }\href {https://doi.org/10.1103/PhysRevLett.25.919} {\bibfield  {journal} {\bibinfo  {journal} {Phys. Rev. Lett.}\ }\textbf {\bibinfo {volume} {25}},\ \bibinfo {pages} {919} (\bibinfo {year} {1970})}\BibitemShut {NoStop}%
\bibitem [{\citenamefont {Su}\ \emph {et~al.}(1979)\citenamefont {Su}, \citenamefont {Schrieffer},\ and\ \citenamefont {Heeger}}]{Su1979solitons}%
  \BibitemOpen
  \bibfield  {author} {\bibinfo {author} {\bibfnamefont {W.~P.}\ \bibnamefont {Su}}, \bibinfo {author} {\bibfnamefont {J.~R.}\ \bibnamefont {Schrieffer}},\ and\ \bibinfo {author} {\bibfnamefont {A.~J.}\ \bibnamefont {Heeger}},\ }\bibfield  {title} {\bibinfo {title} {Solitons in polyacetylene},\ }\href {https://doi.org/10.1103/PhysRevLett.42.1698} {\bibfield  {journal} {\bibinfo  {journal} {Phys. Rev. Lett.}\ }\textbf {\bibinfo {volume} {42}},\ \bibinfo {pages} {1698} (\bibinfo {year} {1979})}\BibitemShut {NoStop}%
\bibitem [{\citenamefont {Holstein}(1959{\natexlab{a}})}]{Holstein1959studiesI}%
  \BibitemOpen
  \bibfield  {author} {\bibinfo {author} {\bibfnamefont {T.}~\bibnamefont {Holstein}},\ }\bibfield  {title} {\bibinfo {title} {Studies of polaron motion: Part {I}. the molecular-crystal model},\ }\href {https://doi.org/https://doi.org/10.1016/0003-4916(59)90002-8} {\bibfield  {journal} {\bibinfo  {journal} {Annals of Physics}\ }\textbf {\bibinfo {volume} {8}},\ \bibinfo {pages} {325} (\bibinfo {year} {1959}{\natexlab{a}})}\BibitemShut {NoStop}%
\bibitem [{\citenamefont {Holstein}(1959{\natexlab{b}})}]{Holstein1959studiesII}%
  \BibitemOpen
  \bibfield  {author} {\bibinfo {author} {\bibfnamefont {T.}~\bibnamefont {Holstein}},\ }\bibfield  {title} {\bibinfo {title} {Studies of polaron motion: Part {II}. the ``small'' polaron},\ }\href {https://doi.org/https://doi.org/10.1016/0003-4916(59)90003-X} {\bibfield  {journal} {\bibinfo  {journal} {Annals of Physics}\ }\textbf {\bibinfo {volume} {8}},\ \bibinfo {pages} {343} (\bibinfo {year} {1959}{\natexlab{b}})}\BibitemShut {NoStop}%
\bibitem [{\citenamefont {Fr{\"o}hlich}(1952)}]{Froelich1952}%
  \BibitemOpen
  \bibfield  {author} {\bibinfo {author} {\bibfnamefont {H.}~\bibnamefont {Fr{\"o}hlich}},\ }\bibfield  {title} {\bibinfo {title} {Interaction of electrons with lattice vibrations},\ }\href {https://doi.org/10.1098/rspa.1952.0212} {\bibfield  {journal} {\bibinfo  {journal} {Proceedings of the Royal Society of London. Series A. Mathematical and Physical Sciences}\ }\textbf {\bibinfo {volume} {215}},\ \bibinfo {pages} {291} (\bibinfo {year} {1952})}\BibitemShut {NoStop}%
\bibitem [{\citenamefont {Sous}\ \emph {et~al.}(2018)\citenamefont {Sous}, \citenamefont {Chakraborty}, \citenamefont {Krems},\ and\ \citenamefont {Berciu}}]{sous2018light}%
  \BibitemOpen
  \bibfield  {author} {\bibinfo {author} {\bibfnamefont {J.}~\bibnamefont {Sous}}, \bibinfo {author} {\bibfnamefont {M.}~\bibnamefont {Chakraborty}}, \bibinfo {author} {\bibfnamefont {R.~V.}\ \bibnamefont {Krems}},\ and\ \bibinfo {author} {\bibfnamefont {M.}~\bibnamefont {Berciu}},\ }\bibfield  {title} {\bibinfo {title} {Light bipolarons stabilized by {P}eierls electron-phonon coupling},\ }\href {https://doi.org/10.1103/PhysRevLett.121.247001} {\bibfield  {journal} {\bibinfo  {journal} {Physical Review Letters}\ }\textbf {\bibinfo {volume} {121}},\ \bibinfo {pages} {247001} (\bibinfo {year} {2018})}\BibitemShut {NoStop}%
\bibitem [{Sup()}]{Supplement}%
  \BibitemOpen
  \href@noop {} {}\bibinfo {note} {See online supplementary materials for additional details on our DMRG and perturbation results, available at \href{\#}{link to be supplied by the publisher}.}\BibitemShut {Stop}%
\bibitem [{\citenamefont {Marchand}\ \emph {et~al.}(2010)\citenamefont {Marchand}, \citenamefont {De~Filippis}, \citenamefont {Cataudella}, \citenamefont {Berciu}, \citenamefont {Nagaosa}, \citenamefont {Prokof'ev}, \citenamefont {Mishchenko},\ and\ \citenamefont {Stamp}}]{Marchand10}%
  \BibitemOpen
  \bibfield  {author} {\bibinfo {author} {\bibfnamefont {D.~J.~J.}\ \bibnamefont {Marchand}}, \bibinfo {author} {\bibfnamefont {G.}~\bibnamefont {De~Filippis}}, \bibinfo {author} {\bibfnamefont {V.}~\bibnamefont {Cataudella}}, \bibinfo {author} {\bibfnamefont {M.}~\bibnamefont {Berciu}}, \bibinfo {author} {\bibfnamefont {N.}~\bibnamefont {Nagaosa}}, \bibinfo {author} {\bibfnamefont {N.~V.}\ \bibnamefont {Prokof'ev}}, \bibinfo {author} {\bibfnamefont {A.~S.}\ \bibnamefont {Mishchenko}},\ and\ \bibinfo {author} {\bibfnamefont {P.~C.~E.}\ \bibnamefont {Stamp}},\ }\bibfield  {title} {\bibinfo {title} {Sharp transition for single polarons in the one-dimensional {S}u-{S}chrieffer-{H}eeger model},\ }\href {https://doi.org/10.1103/PhysRevLett.105.266605} {\bibfield  {journal} {\bibinfo  {journal} {Phys. Rev. Lett.}\ }\textbf {\bibinfo {volume} {105}},\ \bibinfo {pages} {266605} (\bibinfo {year} {2010})}\BibitemShut {NoStop}%
\bibitem [{\citenamefont {G\"otz}\ \emph {et~al.}(2024)\citenamefont {G\"otz}, \citenamefont {Hohenadler},\ and\ \citenamefont {Assaad}}]{Goetz2024phases}%
  \BibitemOpen
  \bibfield  {author} {\bibinfo {author} {\bibfnamefont {A.}~\bibnamefont {G\"otz}}, \bibinfo {author} {\bibfnamefont {M.}~\bibnamefont {Hohenadler}},\ and\ \bibinfo {author} {\bibfnamefont {F.~F.}\ \bibnamefont {Assaad}},\ }\bibfield  {title} {\bibinfo {title} {Phases and exotic phase transitions of a two-dimensional {S}u-{S}chrieffer-{H}eeger model},\ }\href {https://doi.org/10.1103/PhysRevB.109.195154} {\bibfield  {journal} {\bibinfo  {journal} {Phys. Rev. B}\ }\textbf {\bibinfo {volume} {109}},\ \bibinfo {pages} {195154} (\bibinfo {year} {2024})}\BibitemShut {NoStop}%
\bibitem [{\citenamefont {Cai}\ \emph {et~al.}(2021)\citenamefont {Cai}, \citenamefont {Li},\ and\ \citenamefont {Yao}}]{Cai2021antiferromagnetism}%
  \BibitemOpen
  \bibfield  {author} {\bibinfo {author} {\bibfnamefont {X.}~\bibnamefont {Cai}}, \bibinfo {author} {\bibfnamefont {Z.-X.}\ \bibnamefont {Li}},\ and\ \bibinfo {author} {\bibfnamefont {H.}~\bibnamefont {Yao}},\ }\bibfield  {title} {\bibinfo {title} {Antiferromagnetism {I}nduced by bond {S}u-{S}chrieffer-{H}eeger electron-phonon coupling: A quantum {M}onte {C}arlo study},\ }\href {https://doi.org/10.1103/PhysRevLett.127.247203} {\bibfield  {journal} {\bibinfo  {journal} {Phys. Rev. Lett.}\ }\textbf {\bibinfo {volume} {127}},\ \bibinfo {pages} {247203} (\bibinfo {year} {2021})}\BibitemShut {NoStop}%
\bibitem [{\citenamefont {Casebolt}\ \emph {et~al.}(2024)\citenamefont {Casebolt}, \citenamefont {Feng}, \citenamefont {Scalettar}, \citenamefont {Johnston},\ and\ \citenamefont {Batrouni}}]{Casebolt2024magnetic}%
  \BibitemOpen
  \bibfield  {author} {\bibinfo {author} {\bibfnamefont {M.}~\bibnamefont {Casebolt}}, \bibinfo {author} {\bibfnamefont {C.}~\bibnamefont {Feng}}, \bibinfo {author} {\bibfnamefont {R.~T.}\ \bibnamefont {Scalettar}}, \bibinfo {author} {\bibfnamefont {S.}~\bibnamefont {Johnston}},\ and\ \bibinfo {author} {\bibfnamefont {G.~G.}\ \bibnamefont {Batrouni}},\ }\bibfield  {title} {\bibinfo {title} {Magnetic, charge, and bond order in the two-dimensional {S}u-{S}chrieffer-{H}eeger-{H}olstein model},\ }\href {https://doi.org/10.1103/PhysRevB.110.045112} {\bibfield  {journal} {\bibinfo  {journal} {Phys. Rev. B}\ }\textbf {\bibinfo {volume} {110}},\ \bibinfo {pages} {045112} (\bibinfo {year} {2024})}\BibitemShut {NoStop}%
\bibitem [{\citenamefont {Cai}\ \emph {et~al.}(2024)\citenamefont {Cai}, \citenamefont {Han}, \citenamefont {Li}, \citenamefont {Kivelson},\ and\ \citenamefont {Yao}}]{cai2024quantum}%
  \BibitemOpen
  \bibfield  {author} {\bibinfo {author} {\bibfnamefont {X.}~\bibnamefont {Cai}}, \bibinfo {author} {\bibfnamefont {Z.}~\bibnamefont {Han}}, \bibinfo {author} {\bibfnamefont {Z.-X.}\ \bibnamefont {Li}}, \bibinfo {author} {\bibfnamefont {S.~A.}\ \bibnamefont {Kivelson}},\ and\ \bibinfo {author} {\bibfnamefont {H.}~\bibnamefont {Yao}},\ }\bibfield  {title} {\bibinfo {title} {Quantum spin liquid from electron-phonon coupling},\ }\href {https://arxiv.org/abs/2408.04002} {\bibfield  {journal} {\bibinfo  {journal} {arXiv:2408.04002}\ } (\bibinfo {year} {2024})}\BibitemShut {NoStop}%
\bibitem [{\citenamefont {{Malkaruge Costa}}\ \emph {et~al.}(2023)\citenamefont {{Malkaruge Costa}}, \citenamefont {Cohen-Stead}, \citenamefont {{Tanjaroon Ly}}, \citenamefont {Neuhaus},\ and\ \citenamefont {Johnston}}]{MalkarugeCosta2023comparative}%
  \BibitemOpen
  \bibfield  {author} {\bibinfo {author} {\bibfnamefont {S.}~\bibnamefont {{Malkaruge Costa}}}, \bibinfo {author} {\bibfnamefont {B.}~\bibnamefont {Cohen-Stead}}, \bibinfo {author} {\bibfnamefont {A.}~\bibnamefont {{Tanjaroon Ly}}}, \bibinfo {author} {\bibfnamefont {J.}~\bibnamefont {Neuhaus}},\ and\ \bibinfo {author} {\bibfnamefont {S.}~\bibnamefont {Johnston}},\ }\bibfield  {title} {\bibinfo {title} {Comparative determinant quantum {M}onte {C}arlo study of the acoustic and optical variants of the {S}u-{S}chrieffer-{H}eeger model},\ }\href {https://doi.org/10.1103/PhysRevB.108.165138} {\bibfield  {journal} {\bibinfo  {journal} {Phys. Rev. B}\ }\textbf {\bibinfo {volume} {108}},\ \bibinfo {pages} {165138} (\bibinfo {year} {2023})}\BibitemShut {NoStop}%
\bibitem [{\citenamefont {{Tanjaroon Ly}}\ \emph {et~al.}(2023)\citenamefont {{Tanjaroon Ly}}, \citenamefont {Cohen-Stead}, \citenamefont {Malkaruge~Costa},\ and\ \citenamefont {Johnston}}]{TanjaroonLy2023comparative}%
  \BibitemOpen
  \bibfield  {author} {\bibinfo {author} {\bibfnamefont {A.}~\bibnamefont {{Tanjaroon Ly}}}, \bibinfo {author} {\bibfnamefont {B.}~\bibnamefont {Cohen-Stead}}, \bibinfo {author} {\bibfnamefont {S.}~\bibnamefont {Malkaruge~Costa}},\ and\ \bibinfo {author} {\bibfnamefont {S.}~\bibnamefont {Johnston}},\ }\bibfield  {title} {\bibinfo {title} {Comparative study of the superconductivity in the {H}olstein and optical {S}u-{S}chrieffer-{H}eeger models},\ }\href {https://doi.org/10.1103/PhysRevB.108.184501} {\bibfield  {journal} {\bibinfo  {journal} {Phys. Rev. B}\ }\textbf {\bibinfo {volume} {108}},\ \bibinfo {pages} {184501} (\bibinfo {year} {2023})}\BibitemShut {NoStop}%
\bibitem [{\citenamefont {{Tanjaroon Ly}}\ \emph {et~al.}(2025)\citenamefont {{Tanjaroon Ly}}, \citenamefont {Cohen-Stead},\ and\ \citenamefont {Johnston}}]{TanjaroonLy2025antiferromagnetic}%
  \BibitemOpen
  \bibfield  {author} {\bibinfo {author} {\bibfnamefont {A.}~\bibnamefont {{Tanjaroon Ly}}}, \bibinfo {author} {\bibfnamefont {B.}~\bibnamefont {Cohen-Stead}},\ and\ \bibinfo {author} {\bibfnamefont {S.}~\bibnamefont {Johnston}},\ }\bibfield  {title} {\bibinfo {title} {Antiferromagnetic and bond-order-wave phases in the half-filled two-dimensional optical {S}u-{S}chrieffer-{H}eeger-{H}ubbard model},\ }\href {https://arxiv.org/abs/2502.14196} {\bibfield  {journal} {\bibinfo  {journal} {arXiv:2502.14196}\ } (\bibinfo {year} {2025})}\BibitemShut {NoStop}%
\bibitem [{\citenamefont {Cai}\ \emph {et~al.}(2023)\citenamefont {Cai}, \citenamefont {Li},\ and\ \citenamefont {Yao}}]{cai2023hightemperature}%
  \BibitemOpen
  \bibfield  {author} {\bibinfo {author} {\bibfnamefont {X.}~\bibnamefont {Cai}}, \bibinfo {author} {\bibfnamefont {Z.-X.}\ \bibnamefont {Li}},\ and\ \bibinfo {author} {\bibfnamefont {H.}~\bibnamefont {Yao}},\ }\bibfield  {title} {\bibinfo {title} {High-temperature superconductivity induced by the {S}u-{S}chrieffer-{H}eeger electron-phonon coupling},\ }\href {https://arxiv.org/abs/2308.06222} {\bibfield  {journal} {\bibinfo  {journal} {arXiv:2308.06222}\ } (\bibinfo {year} {2023})}\BibitemShut {NoStop}%
\bibitem [{\citenamefont {Capone}\ \emph {et~al.}(1997)\citenamefont {Capone}, \citenamefont {Stephan},\ and\ \citenamefont {Grilli}}]{Capone1997small}%
  \BibitemOpen
  \bibfield  {author} {\bibinfo {author} {\bibfnamefont {M.}~\bibnamefont {Capone}}, \bibinfo {author} {\bibfnamefont {W.}~\bibnamefont {Stephan}},\ and\ \bibinfo {author} {\bibfnamefont {M.}~\bibnamefont {Grilli}},\ }\bibfield  {title} {\bibinfo {title} {Small-polaron formation and optical absorption in {Su-Schrieffer-Heeger} and {H}olstein models},\ }\href {https://doi.org/10.1103/PhysRevB.56.4484} {\bibfield  {journal} {\bibinfo  {journal} {Phys. Rev. B}\ }\textbf {\bibinfo {volume} {56}},\ \bibinfo {pages} {4484} (\bibinfo {year} {1997})}\BibitemShut {NoStop}%
\bibitem [{\citenamefont {Banerjee}\ \emph {et~al.}(2023)\citenamefont {Banerjee}, \citenamefont {Thomas}, \citenamefont {Nocera},\ and\ \citenamefont {Johnston}}]{Banerjee2023groundstate}%
  \BibitemOpen
  \bibfield  {author} {\bibinfo {author} {\bibfnamefont {D.}~\bibnamefont {Banerjee}}, \bibinfo {author} {\bibfnamefont {J.}~\bibnamefont {Thomas}}, \bibinfo {author} {\bibfnamefont {A.}~\bibnamefont {Nocera}},\ and\ \bibinfo {author} {\bibfnamefont {S.}~\bibnamefont {Johnston}},\ }\bibfield  {title} {\bibinfo {title} {Ground-state and spectral properties of the doped one-dimensional optical {H}ubbard-{S}u-{S}chrieffer-{H}eeger model},\ }\href {https://doi.org/10.1103/PhysRevB.107.235113} {\bibfield  {journal} {\bibinfo  {journal} {Phys. Rev. B}\ }\textbf {\bibinfo {volume} {107}},\ \bibinfo {pages} {235113} (\bibinfo {year} {2023})}\BibitemShut {NoStop}%
\bibitem [{\citenamefont {White}(1992)}]{DMRG_Steve_white_PRL}%
  \BibitemOpen
  \bibfield  {author} {\bibinfo {author} {\bibfnamefont {S.~R.}\ \bibnamefont {White}},\ }\bibfield  {title} {\bibinfo {title} {Density matrix formulation for quantum renormalization groups},\ }\href {https://doi.org/10.1103/PhysRevLett.69.2863} {\bibfield  {journal} {\bibinfo  {journal} {Phys. Rev. Lett.}\ }\textbf {\bibinfo {volume} {69}},\ \bibinfo {pages} {2863} (\bibinfo {year} {1992})}\BibitemShut {NoStop}%
\bibitem [{\citenamefont {Alvarez}(2009)}]{DMRGpp}%
  \BibitemOpen
  \bibfield  {author} {\bibinfo {author} {\bibfnamefont {G.}~\bibnamefont {Alvarez}},\ }\bibfield  {title} {\bibinfo {title} {The density matrix renormalization group for strongly correlated electron systems: A generic implementation},\ }\href {https://doi.org/https://doi.org/10.1016/j.cpc.2009.02.016} {\bibfield  {journal} {\bibinfo  {journal} {Computer Physics Communications}\ }\textbf {\bibinfo {volume} {180}},\ \bibinfo {pages} {1572} (\bibinfo {year} {2009})}\BibitemShut {NoStop}%
\bibitem [{\citenamefont {Damascelli}\ \emph {et~al.}(2003)\citenamefont {Damascelli}, \citenamefont {Hussain},\ and\ \citenamefont {Shen}}]{damascelli2003angle}%
  \BibitemOpen
  \bibfield  {author} {\bibinfo {author} {\bibfnamefont {A.}~\bibnamefont {Damascelli}}, \bibinfo {author} {\bibfnamefont {Z.}~\bibnamefont {Hussain}},\ and\ \bibinfo {author} {\bibfnamefont {Z.-X.}\ \bibnamefont {Shen}},\ }\bibfield  {title} {\bibinfo {title} {Angle-resolved photoemission studies of the cuprate superconductors},\ }\href {https://doi.org/https://doi.org/10.1103/RevModPhys.75.473} {\bibfield  {journal} {\bibinfo  {journal} {Reviews of modern physics}\ }\textbf {\bibinfo {volume} {75}},\ \bibinfo {pages} {473} (\bibinfo {year} {2003})}\BibitemShut {NoStop}%
\bibitem [{\citenamefont {Sobota}\ \emph {et~al.}(2021)\citenamefont {Sobota}, \citenamefont {He},\ and\ \citenamefont {Shen}}]{sobota2021angle}%
  \BibitemOpen
  \bibfield  {author} {\bibinfo {author} {\bibfnamefont {J.~A.}\ \bibnamefont {Sobota}}, \bibinfo {author} {\bibfnamefont {Y.}~\bibnamefont {He}},\ and\ \bibinfo {author} {\bibfnamefont {Z.-X.}\ \bibnamefont {Shen}},\ }\bibfield  {title} {\bibinfo {title} {Angle-resolved photoemission studies of quantum materials},\ }\href {https://doi.org/https://doi.org/10.1103/RevModPhys.93.025006} {\bibfield  {journal} {\bibinfo  {journal} {Reviews of Modern Physics}\ }\textbf {\bibinfo {volume} {93}},\ \bibinfo {pages} {025006} (\bibinfo {year} {2021})}\BibitemShut {NoStop}%
\bibitem [{\citenamefont {Nocera}\ and\ \citenamefont {Alvarez}(2016)}]{Nocera2016spectralfunction}%
  \BibitemOpen
  \bibfield  {author} {\bibinfo {author} {\bibfnamefont {A.}~\bibnamefont {Nocera}}\ and\ \bibinfo {author} {\bibfnamefont {G.}~\bibnamefont {Alvarez}},\ }\bibfield  {title} {\bibinfo {title} {Spectral functions with the density matrix renormalization group: {K}rylov-space approach for correction vectors},\ }\href {https://doi.org/10.1103/PhysRevE.94.053308} {\bibfield  {journal} {\bibinfo  {journal} {Phys. Rev. E}\ }\textbf {\bibinfo {volume} {94}},\ \bibinfo {pages} {053308} (\bibinfo {year} {2016})}\BibitemShut {NoStop}%
\bibitem [{\citenamefont {Engelsberg}\ and\ \citenamefont {Schrieffer}(1963)}]{Engelsberg1963coupled}%
  \BibitemOpen
  \bibfield  {author} {\bibinfo {author} {\bibfnamefont {S.}~\bibnamefont {Engelsberg}}\ and\ \bibinfo {author} {\bibfnamefont {J.~R.}\ \bibnamefont {Schrieffer}},\ }\bibfield  {title} {\bibinfo {title} {Coupled electron-phonon system},\ }\href {https://doi.org/10.1103/PhysRev.131.993} {\bibfield  {journal} {\bibinfo  {journal} {Phys. Rev.}\ }\textbf {\bibinfo {volume} {131}},\ \bibinfo {pages} {993} (\bibinfo {year} {1963})}\BibitemShut {NoStop}%
\bibitem [{\citenamefont {Bon\ifmmode~\check{c}\else \v{c}\fi{}a}\ \emph {et~al.}(2019)\citenamefont {Bon\ifmmode~\check{c}\else \v{c}\fi{}a}, \citenamefont {Trugman},\ and\ \citenamefont {Berciu}}]{Bonca2019spectral}%
  \BibitemOpen
  \bibfield  {author} {\bibinfo {author} {\bibfnamefont {J.}~\bibnamefont {Bon\ifmmode~\check{c}\else \v{c}\fi{}a}}, \bibinfo {author} {\bibfnamefont {S.~A.}\ \bibnamefont {Trugman}},\ and\ \bibinfo {author} {\bibfnamefont {M.}~\bibnamefont {Berciu}},\ }\bibfield  {title} {\bibinfo {title} {Spectral function of the {H}olstein polaron at finite temperature},\ }\href {https://doi.org/10.1103/PhysRevB.100.094307} {\bibfield  {journal} {\bibinfo  {journal} {Phys. Rev. B}\ }\textbf {\bibinfo {volume} {100}},\ \bibinfo {pages} {094307} (\bibinfo {year} {2019})}\BibitemShut {NoStop}%
\bibitem [{\citenamefont {Hague}\ \emph {et~al.}(2006)\citenamefont {Hague}, \citenamefont {Kornilovitch}, \citenamefont {Alexandrov},\ and\ \citenamefont {Samson}}]{hague2006effects}%
  \BibitemOpen
  \bibfield  {author} {\bibinfo {author} {\bibfnamefont {J.~P.}\ \bibnamefont {Hague}}, \bibinfo {author} {\bibfnamefont {P.~E.}\ \bibnamefont {Kornilovitch}}, \bibinfo {author} {\bibfnamefont {A.~S.}\ \bibnamefont {Alexandrov}},\ and\ \bibinfo {author} {\bibfnamefont {J.~H.}\ \bibnamefont {Samson}},\ }\bibfield  {title} {\bibinfo {title} {Effects of lattice geometry and interaction range on polaron dynamics},\ }\href {https://doi.org/10.1103/PhysRevB.73.054303} {\bibfield  {journal} {\bibinfo  {journal} {Phys. Rev. B}\ }\textbf {\bibinfo {volume} {73}},\ \bibinfo {pages} {054303} (\bibinfo {year} {2006})}\BibitemShut {NoStop}%
\bibitem [{\citenamefont {Goodvin}\ \emph {et~al.}(2006)\citenamefont {Goodvin}, \citenamefont {Berciu},\ and\ \citenamefont {Sawatzky}}]{goodvin2006green}%
  \BibitemOpen
  \bibfield  {author} {\bibinfo {author} {\bibfnamefont {G.~L.}\ \bibnamefont {Goodvin}}, \bibinfo {author} {\bibfnamefont {M.}~\bibnamefont {Berciu}},\ and\ \bibinfo {author} {\bibfnamefont {G.~A.}\ \bibnamefont {Sawatzky}},\ }\bibfield  {title} {\bibinfo {title} {Green's function of the {H}olstein polaron},\ }\href {https://doi.org/10.1103/PhysRevB.74.245104} {\bibfield  {journal} {\bibinfo  {journal} {Phys. Rev. B}\ }\textbf {\bibinfo {volume} {74}},\ \bibinfo {pages} {245104} (\bibinfo {year} {2006})}\BibitemShut {NoStop}%
\bibitem [{\citenamefont {Bon{\v c}a}\ \emph {et~al.}(2000)\citenamefont {Bon{\v c}a}, \citenamefont {Katra{\v s}nik},\ and\ \citenamefont {Trugman}}]{bonca2000mobile}%
  \BibitemOpen
  \bibfield  {author} {\bibinfo {author} {\bibfnamefont {J.}~\bibnamefont {Bon{\v c}a}}, \bibinfo {author} {\bibfnamefont {T.}~\bibnamefont {Katra{\v s}nik}},\ and\ \bibinfo {author} {\bibfnamefont {S.~A.}\ \bibnamefont {Trugman}},\ }\bibfield  {title} {\bibinfo {title} {Mobile bipolaron},\ }\href {https://doi.org/10.1103/PhysRevLett.84.3153} {\bibfield  {journal} {\bibinfo  {journal} {Phys. Rev. Lett.}\ }\textbf {\bibinfo {volume} {84}},\ \bibinfo {pages} {3153} (\bibinfo {year} {2000})}\BibitemShut {NoStop}%
\bibitem [{\citenamefont {Sous}\ \emph {et~al.}(2017)\citenamefont {Sous}, \citenamefont {Chakraborty}, \citenamefont {Adolphs}, \citenamefont {Krems},\ and\ \citenamefont {Berciu}}]{sous2017phonon}%
  \BibitemOpen
  \bibfield  {author} {\bibinfo {author} {\bibfnamefont {J.}~\bibnamefont {Sous}}, \bibinfo {author} {\bibfnamefont {M.}~\bibnamefont {Chakraborty}}, \bibinfo {author} {\bibfnamefont {C.~P.~J.}\ \bibnamefont {Adolphs}}, \bibinfo {author} {\bibfnamefont {R.}~\bibnamefont {Krems}},\ and\ \bibinfo {author} {\bibfnamefont {M.}~\bibnamefont {Berciu}},\ }\bibfield  {title} {\bibinfo {title} {Phonon-mediated repulsion, sharp transitions and (quasi) self-trapping in the extended {P}eierls-{H}ubbard model},\ }\href {https://doi.org/https://doi.org/10.1038/s41598-017-01228-y} {\bibfield  {journal} {\bibinfo  {journal} {Scientific reports}\ }\textbf {\bibinfo {volume} {7}},\ \bibinfo {pages} {1169} (\bibinfo {year} {2017})}\BibitemShut {NoStop}%
\bibitem [{\citenamefont {Nocera}\ \emph {et~al.}(2021)\citenamefont {Nocera}, \citenamefont {Sous}, \citenamefont {Feiguin},\ and\ \citenamefont {Berciu}}]{nocera2021bipolaron}%
  \BibitemOpen
  \bibfield  {author} {\bibinfo {author} {\bibfnamefont {A.}~\bibnamefont {Nocera}}, \bibinfo {author} {\bibfnamefont {J.}~\bibnamefont {Sous}}, \bibinfo {author} {\bibfnamefont {A.~E.}\ \bibnamefont {Feiguin}},\ and\ \bibinfo {author} {\bibfnamefont {M.}~\bibnamefont {Berciu}},\ }\bibfield  {title} {\bibinfo {title} {Bipolaron liquids at strong {P}eierls electron-phonon couplings},\ }\href {https://doi.org/10.1103/PhysRevB.104.L201109} {\bibfield  {journal} {\bibinfo  {journal} {Phys. Rev. B}\ }\textbf {\bibinfo {volume} {104}},\ \bibinfo {pages} {L201109} (\bibinfo {year} {2021})}\BibitemShut {NoStop}%
\bibitem [{\citenamefont {Takahashi}(1977)}]{Takahashi77}%
  \BibitemOpen
  \bibfield  {author} {\bibinfo {author} {\bibfnamefont {M.}~\bibnamefont {Takahashi}},\ }\bibfield  {title} {\bibinfo {title} {Half-filled {H}ubbard model at low temperature},\ }\href {https://doi.org/10.1088/0022-3719/10/8/031} {\bibfield  {journal} {\bibinfo  {journal} {Journal of Physics C: Solid State Physics}\ }\textbf {\bibinfo {volume} {10}},\ \bibinfo {pages} {1289} (\bibinfo {year} {1977})}\BibitemShut {NoStop}%
\bibitem [{\citenamefont {Chen}\ \emph {et~al.}(2021)\citenamefont {Chen}, \citenamefont {Wang}, \citenamefont {Rebec}, \citenamefont {Jia}, \citenamefont {Hashimoto}, \citenamefont {Lu}, \citenamefont {Moritz}, \citenamefont {Moore}, \citenamefont {Devereaux},\ and\ \citenamefont {Shen}}]{Chen2021anomalously}%
  \BibitemOpen
  \bibfield  {author} {\bibinfo {author} {\bibfnamefont {Z.}~\bibnamefont {Chen}}, \bibinfo {author} {\bibfnamefont {Y.}~\bibnamefont {Wang}}, \bibinfo {author} {\bibfnamefont {S.~N.}\ \bibnamefont {Rebec}}, \bibinfo {author} {\bibfnamefont {T.}~\bibnamefont {Jia}}, \bibinfo {author} {\bibfnamefont {M.}~\bibnamefont {Hashimoto}}, \bibinfo {author} {\bibfnamefont {D.}~\bibnamefont {Lu}}, \bibinfo {author} {\bibfnamefont {B.}~\bibnamefont {Moritz}}, \bibinfo {author} {\bibfnamefont {R.~G.}\ \bibnamefont {Moore}}, \bibinfo {author} {\bibfnamefont {T.~P.}\ \bibnamefont {Devereaux}},\ and\ \bibinfo {author} {\bibfnamefont {Z.-X.}\ \bibnamefont {Shen}},\ }\bibfield  {title} {\bibinfo {title} {Anomalously strong near-neighbor attraction in doped {1D} cuprate chains},\ }\href {https://doi.org/10.1126/science.abf5174} {\bibfield  {journal} {\bibinfo  {journal} {Science}\ }\textbf {\bibinfo {volume} {373}},\ \bibinfo {pages} {1235} (\bibinfo {year} {2021})}\BibitemShut {NoStop}%
\end{thebibliography}%


%
\end{document}


\preprint{}
\title{Supplementary Material for ``Spectral signatures of residual electron pairing in the extended-{H}ubbard--{S}u-{S}chrieffer-{H}eeger model''}

\author{Debshikha Banerjee\orcidlink{0009-0001-2925-9724}}
\affiliation{Department of Physics and Astronomy, The University of Tennessee, Knoxville, Tennessee 37996, USA}
\affiliation{Institute for Advanced Materials and Manufacturing, University of Tennessee, Knoxville, Tennessee 37996, USA\looseness=-1}

\author{Alberto Nocera\orcidlink{0000-0001-9722-6388}}
\affiliation{Department of Physics Astronomy, University of British Columbia, Vancouver, British Columbia, Canada V6T 1Z1}
\affiliation{Stewart Blusson Quantum Matter Institute, University of British Columbia, Vancouver, British Columbia, Canada
V6T 1Z4}

\author{George A. Sawatzky\orcidlink{0000-0003-1265-2770}}
\affiliation{Department of Physics Astronomy, University of British Columbia, Vancouver, British Columbia, Canada V6T 1Z1}
\affiliation{Stewart Blusson Quantum Matter Institute, University of British Columbia, Vancouver, British Columbia, Canada
V6T 1Z4}

\author{Mona Berciu\orcidlink{0000-0002-6736-1893
}}
\affiliation{Department of Physics Astronomy, University of British Columbia, Vancouver, British Columbia, Canada V6T 1Z1}
\affiliation{Stewart Blusson Quantum Matter Institute, University of British Columbia, Vancouver, British Columbia, Canada
V6T 1Z4}

\author{Steven Johnston\orcidlink{0000-0002-2343-0113}}
\affiliation{Department of Physics and Astronomy, The University of Tennessee, Knoxville, Tennessee 37996, USA}
\affiliation{Institute for Advanced Materials and Manufacturing, University of Tennessee, Knoxville, Tennessee 37996, USA\looseness=-1}

\date{\today}

\maketitle

\section{The effective Hamiltonian in the anti-adiabatic limit}
This section presents a detailed perturbation analysis of our model in the anti-adiabatic limit where $\Omega$ and $U$ are much larger than all other energy scales. 

We partition our Hamiltonian as ${\cal H}={\cal H}_0 + T + V$, where
\begin{equation}
  \label{1}
{\cal H}_0 = \Omega \sum_i b^\dagger_i b^{\phantom\dagger}_i + U \sum_i n_{i\uparrow} n_{i\downarrow}
\end{equation}
are the terms that are diagonal in the real space representation,  
\begin{equation}
    T = - t\sum_{i,\sigma} [c^\dagger_{i\sigma} c^{\phantom\dagger}_{i+1,\sigma} + \text{H.c.}]
\end{equation}
is the hopping, and 
\begin{equation}
  V = g \sum_{i\sigma} [c^\dagger_{i\sigma} c^{\phantom\dagger}_{i+1,\sigma} + \text{H.c.}]( b^\dagger_{i+1}+b^{\phantom\dagger}_{i+1} -  b^\dagger_i -b^{\phantom\dagger}_i)  
\end{equation}
is the Peierls/SSH electron-phonon coupling.

In the antiadiabatic limit we generalize Takahashi's method \cite{Takahashi77} to get an effective Hamiltonian for the low-lying eigenstates using perturbation theory. Specifically, for configurations with two electrons in the system, we define projection operators $P_1$ and $P_2$ that project onto different manifolds of zero-phonon configurations. 
Let $P_1$ project onto the manifold of zero-phonon configurations $c^\dagger_{i\uparrow} c^\dagger_{j\downarrow}|0\rangle$ with $i\ne j$, which are eigenstates of ${\cal H}_0$ with energy $E_1=0$; and let $P_2$ project onto the manifold of zero-phonon configurations $c^\dagger_{i\uparrow} c^\dagger_{i\downarrow}|0\rangle$, which are eigenstates of ${\cal H}_0$ with energy $E_2=U$. To second-order perturbation theory in $t$ and $g$, the effective Hamiltonian in the zero-phonon manifold is
\begin{equation}
  \label{2}
{\cal H}_{\rm eff} = {\cal H}_0 + P_0 (T+V) P_0 + \sum_{i,j=1,2} P_i (T+V) {S_i + S_j\over 2} (T+V) P_j, 
\end{equation}  
where $P_0=P_1+P_2$ is the total projector onto the zero-phonon manifold, and $S_i = \frac{1-P_0}{E_i -{\cal H}_0}$ are related to projectors onto the many-phonon manifolds.

Calculating these contributions is a straightforward task. To simplify the notation, we note that because ${\cal H}$ is invariant to translations, it is convenient to work with states with a fixed total momentum $k$. Specifically, in the singlet sector, we define
\begin{equation}
  \label{3}
|k,0\rangle = \sum_i {e^{\mathrm{i} k R_i}\over \sqrt{N}}  c^\dagger_{i\uparrow} c^\dagger_{i\downarrow}|0\rangle
\end{equation}  
and for all $n\ge 1$:
\begin{equation}\label{4}
|k,n\rangle = \frac{1}{\sqrt{2N}}\sum_i e^{\mathrm{i} k (R_i+{na\over 2})} \left(c^\dagger_{i\uparrow} c^\dagger_{i+n,\downarrow}-c^\dagger_{i+n,\uparrow}c^\dagger_{i\downarrow }\right)|0\rangle, 
\end{equation} 
so that $P_1= \sum_{k, n\ge 1} |k,n\rangle\langle k,n|$,  $P_2=\sum_k |k,0\rangle\langle k, 0|$. The sums over $k$ are over the $N$ allowed $k$ values inside the first Brillouin zone $ka \in (-\pi, \pi]$, where $n\rightarrow \infty$ is the number of sites.  

Straightforward calculations give the expressions of  ${\cal H}_{\rm eff} |k,n\rangle$ for any $n\ge 0$. For example, $ {\cal H}_{\rm eff} |k,0\rangle = U_0(k) |k,0\rangle - \sqrt{2} t_1(k) |k,1\rangle+ {\tilde t}_2(k) |k,2\rangle $, where
\begin{align}
 U_0(k)&=  U- {8g^2\over \Omega}[1+\cos(ka)], \nonumber \\
 U_1(k)&= - \left({g^2\over \Omega} +{2g^2\over \Omega+U}  \right)[4-2\cos(ka)], \label{5}\\
 U_{n\ge 2}(k)&= -{8g^2\over \Omega},\nonumber
\end{align}
and $t_1(k)=2t \cos{\tfrac{ka}{2}}$, $t_2(k) = {2g^2\over \Omega}\cos(ka)$, and ${\tilde t}_2(k) = \sqrt{2} \left({g^2\over \Omega} +{g^2\over \Omega-U}   \right)[1+\cos(ka)]$. Here, $U_n(k)$ characterizes the effective interaction between particles when they are $n$-sites apart. The parts that are $k$-independent (like $U$ and $-{8g^2\over \Omega}$, which is twice the polaron deformation energy within this approximation) signal density-density-type of interactions. The $k$-dependent parts come from pair-hopping type of interactions, which are mediated by phonon exchange between the two moving carriers, and which have already been highlighted in Ref. ~\cite{sous2018light}.

Hopping links a state $|k, n\rangle$ to $|k, n\pm 1\rangle$, resulting in the $t_1(k)$ matrix element proportional to $t$. Effective 2nd nearest neighbor hopping, described by $t_2(k)$ matrix elements, is dynamically generated through phonon emission and absorption by the same carrier, typical for SSH coupling \cite{Marchand10}. Finally, ${\tilde t}_2(k)-t_2(k)$ describes an effective phonon-mediated interaction whereby phonon exchange moves two initially on-site carriers two sites apart, and vice-versa~\cite{sous2018light}.

Next, we define the singlet momentum states with total momentum $k$ and relative momentum $2q \ge 0$ as
\begin{equation}
  \label{6}
|k, q\rangle = \frac{1}{\sqrt{2(1+\delta_{q,0})}}\left(c^\dagger_{{k\over 2}+q, \uparrow}c^\dagger_{{k\over 2}-q, \downarrow}-c^\dagger_{{k\over 2}-q, \uparrow} c^\dagger_{{k\over 2}+q, \downarrow}\right)|0\rangle
\end{equation}  
In this basis, we find that:
\begin{multline}
  \label{8}
        {\cal H}_{\rm eff} =\sum_{k}\left[\sum_{q\ge 0} \epsilon(k,Q) |k,q\rangle \langle k, q| \right.\\
     \left.    + {2\over N} \sum_{q,q'\ge 0} \left({u_0(k)\over 2} + 2{u}_1(k)\cos(qa)\cos(q'a) + u_2(k) [\cos(2qa) + \cos(2q'a)] \right)|k,q\rangle\langle k,q'|  \right]
\end{multline}

Here, $\epsilon(k,q)= E_P({k\over 2} + q) + E_P({k\over 2} - q)  $ is the total energy of two non-interacting (far apart) polarons, where the polaron energy is (within this approximation) $E_P(k) = -{4g^2\over \Omega}- 2t \cos(ka) + {2g^2\over \Omega} \cos(2ka)$. Interactions, whether due to the on-site repulsion $U$ or to the phonon-mediated terms discussed above, are described by the terms on the second line,  and result in momentum exchange between the two polarons. For convenience, we introduced $u_0(k) = U_0(k)+{8g^2\over \Omega}$, $u_1(k)= U_1(k) -t_2(k)+{8g^2\over \Omega}$ and $u_2(k) = \sqrt{2} {\tilde t}_2(k) - 2t_2(k)$.

\section{Two-particle propagator}
Next, we calculate the propagator
\begin{equation}
  \label{9}
G(k,z;Q,q) = \langle k, Q|{\hat G}_{\rm eff}(z)|k,q\rangle
\end{equation}
where $z=\omega+i\eta$ is the energy including a small broadening $\eta>0$, and ${\hat G}_{\rm eff}(z)=[ z-  {\cal H}_{\rm eff}]^{-1}$. 

We use Dyson's identity ${\hat G}_{\rm eff}(z) = {\hat G}_{0}(z) + {\hat G}_{\rm eff}(z) {\hat U}_{\rm eff} {\hat G}_{0}(z)$ if $ {\cal H}_{\rm eff} = H_0 + {\hat U}_{\rm eff}$, where we identify 
\begin{equation}
H_0 = \sum_{k}\sum_{q\ge 0} \epsilon(k,q) |k,q\rangle \langle k, q|
\end{equation}
as the non-interacting part of ${\cal H}_{\rm eff}$. Using $ \langle k, Q|{\hat G}_{0}(z)|k,q\rangle= \delta_{q,Q}G_0(k,z; q)$, where  $ G_0(k,z; q)=[z-\epsilon(k,q)]^{-1}$, we find
\begin{align}
      \nonumber
      G(k,z;Q,q) = G_0(k,z;q)\big\{\delta_{q,Q}&+ 2[ u_0(k) +u_2(k) \cos(2qa)] A_0(k,z;Q) \\
      &+ 4 {u}_1(k) \cos (qa) A_1(k,z;Q) + 2 u_2(k) A_2(k,z;Q)\big\} 
      \label{9b}
\end{align}
where $A_n(k,z;Q) = {1\over N} \sum_{q\ge0} \cos(nqa) G(k,z;Q,q) $ for $n=0,1,2$. Scaling  $A_n(k,z;Q) =  { G_0(k,z;Q) \over N}  a_n(k,z;q)$ and combining with Eq. (\ref{9b}), we obtain the linear system of three equations:
\begin{align} \nonumber
  a_0 &=  1 + \left[2 u_0(k) g_0(k,z) + 2 u_2(k)g_2(k,z)\right]a_0 + 4 {u}_1(k) g_1(k,z) a_1 + 2 u_2(k)g_0(k,z) a_2,\\
  a_1 &=  \cos(Qa) + \left[2 u_0(k) g_1(k,z) + u_2(k)(g_1(k,z)+g_3(k,z))\right]a_0 + 2 {u}_1(k) [g_0(k,z)+g_2(k,z)])a_1 +  2u_2(k)g_1(k,z) a_2, \nonumber\\
  a_2 &=  \cos(2Qa) + \left[2 U_0(k) g_2(k,z) + u_2(k)(g_0(k,z)+g_4(k,z))\right]a_0 + 2 {u}_1(k) [g_1(k,z)+g_3(k,z)]a_1 +  2u_2(k)g_1(k,z) a_2,
\end{align}
where we use $a_n\equiv a_n(k,z;Q)$ for brevity and we defined $g_n(k,z)= {1\over N} \sum_{q\ge0} \cos(nqa) G_0(k,z;q)$. These latter real-space propagators can be calculated numerically.

Solving this system numerically, we find the corresponding $a_n(k,z;Q)$, $n=0,1,2$ which can then be used in Eq. (\ref{9b}) to calculate any $G(k,z;Q,q)$.

\section{Comparison with DMRG results}
The DMRG spectral weight is for adding a particle with momentum $k$ and spin-$\downarrow$ to the ground-state for 1 particle with spin-$\uparrow$, i.e. to a spin-up polaron ground-state. In our approximation, this maps onto $\langle0| c_{0,\uparrow} c_{k\downarrow} {\hat G}_{\rm eff}(z) c^\dagger_{k\downarrow} c^\dagger_{0\uparrow}|0\rangle= G(k,z;{k\over2}, {k\over2})$, because after the projection on the zero-phonon states, $c^\dagger_{0\uparrow}|0\rangle$ describes a polaron with GS momentum $k=0$ and energy $E_P(k=0)$. According to the calculation described above, the propagator in the singlet sector corresponds to $q=Q=k/2$ and is, thus:
\begin{multline}
  \label{11}
  G(k,z;{\tfrac{k}{2}}, {\tfrac{k}{2}}) = G_0(k,z;{\tfrac{k}{2}}) \\
  + {2\over N} [G_0(k,z;\tfrac{k}{2})]^2 \left(2\left[ u_0(k) +u_2(k) \cos(ka)\right]a_0(k,z;\tfrac{k}{2}) + 4 u_1(k) \cos{\tfrac{k}{2}} a_1(k,z;\tfrac{k}{2}) + 2 u_2(k) a_2(k,z;\tfrac{k}{2}) \right)
\end{multline}

The first line is the expected result if there are no interactions between the two polarons. As already discussed, it has poles at the energy $\omega = E_P(0)+E_P(k)$, which is the expected low energy feature in the absence of interactions. The second line shows the effect of the interactions. In particular, the existence of poles in the long-expression multiplying $[G_0(k,z;{k\over2}) ]^2 $ indicates the appearance of bound two-particle (bipolaron) eigenstates.
\begin{figure}
    \centering
    \includegraphics[width=\columnwidth]{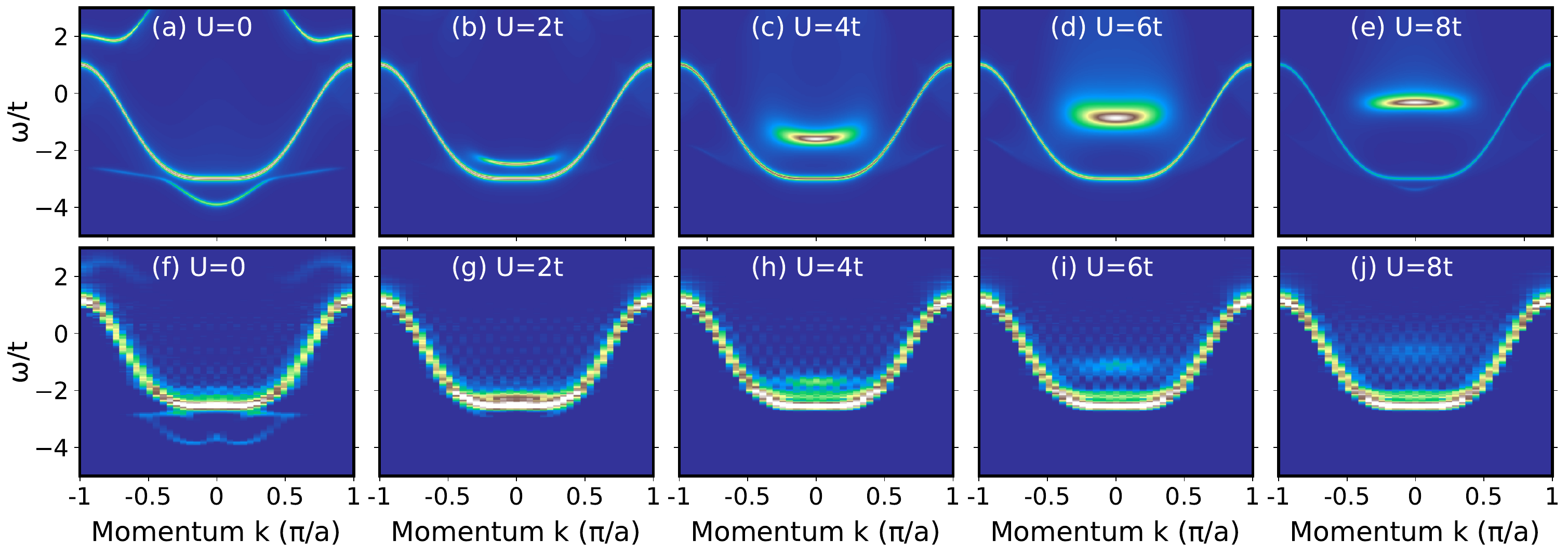}
    \caption{Comparison of the electron addition spectral function $A^+(k,\omega)$ computed using second-order perturbation theory (panels a-e) and DMRG (panels f-j) for the Hubbard-SSH model. The model parameters used here are $\Omega=10t$, $V = 0$, and $U$ as indicated in each panel. For panels (a-e) electron-phonon coupling is fixed at $g=1.58\: (\lambda=0.125)$ and for panels (f-j) it is $g=1.8 \: (\lambda=0.163)$. DMRG results are computed for an $L=40$ site chain.
    }
    \label{fig:PT_vs_DMRG}
\end{figure}
To make this more apparent, consider first the case without electron-phonon coupling, $g=0$, in which case the calculation above is exact and leads to the two-particle propagator for a pure Hubbard model:
\begin{equation}
  \label{12}
 G(k,z;{\tfrac{k}{2}}, {\tfrac{k}{2}}) = {1\over z + 4t \cos^2 {\tfrac{ka}{2}}}+ {1\over N} \left[{1\over z + 4t \cos^2 {\tfrac{ka}{2}}}\right]^2 \frac{U}{1- \frac{U}{\sqrt{z-4t\cos{\tfrac{ka}{2}}} \sqrt{z+4t\cos{\tfrac{ka}{2}}}}}
\end{equation}
This has a structure similar to Eq. (\ref{11}). In the absence of interactions (first term), the poles appear at $\omega= - 4t \cos^2 {\tfrac{ka}{2}}= \epsilon(0)+\epsilon(k)$, where $\epsilon(k)=-2t \cos(ka)$ is the energy of a free carrier with momentum $k$. The second term appears because of the on-site interaction $U$, and allows for a new pole when $U = \sqrt{z-4t\cos{ka\over2}} \sqrt{z+4t\cos{ka\over2}}$. A discrete solution is only possible outside the continuum $ - 4t \cos{ka\over2}\le\omega\le 4t \cos{ka\over2}$ which describes the convolution of two single carrier spectra with total momentum $k$, ie the continuum of states accessible due to carrier-carrier scattering. If $U>0$, there is a new pole above this continuum at $E_B(k) = \sqrt{U^2 + 16 t^2 \cos^2{ka\over2}}$ describing an 'anti-bound' state due to strong on-site repulsion between the two carriers, when they are at the same site. For $U<0$, the new pole appears below the continuum, at $E_B(k) =- \sqrt{U^2 + 16 t^2 \cos^2{ka\over2}}$  and describes a pair bound by the on-site attraction \cite{Sawatzky77}.  Note that the weight of this second term scales like $1/N$. This is expected because this term is proportional to the probability of the two carriers being at the same site and thus to interact, and it means that in the thermodynamic limit, this bound state becomes invisible in this representation (one can see it if the propagator is projected on the $|k,0\rangle$ state of Eq. (3)). In a system with a finite but low concentration of carriers, one would expect that this term scales like the carrier concentration $x$. In the results shown below, we set (rather arbitrarily) this prefactor $1/N$ to a finite value such that the two terms have similar spectral weight, so that they are both visible in the same color plot. Note, however, that when we calculate the various propagators we use the thermodynamic limit.

We expect similar behavior in the presence of electron-phonon coupling, at finite $g$. In this case, the effective on-site interaction described by $u_0(k)=U- {8g^2\over \Omega} \cos(ka)$ can become attractive near $k=0$ for a strong enough electron-phonon coupling, promoting the formation of bipolarons. However, the Peierls coupling also mediates effective interactions between nearest-neighbor (nn) and second nearest-neighbor (nnn) carriers, described by $u_1(k), u_2(k)$, which do not depend on $U$. It is thus possible that for a large enough $U$, $u_0(k)>0$ and there are no stable bipolarons, however, the nn and nnn interactions $u_1(k),u_2(k)$ are still attractive at small enough $|k|$ and can mediate the appearance of a 'resonance'. One way to think about this is as a quasibound-state (primarily on nn sites) with a finite lifetime (meaning, a broadening much larger than $\eta$) because it lies at higher energies and it is energetically more convenient to unbind eventually into two polarons.

We validated this interpretation by comparing results obtained in using this approximation with DMRG results for $\Omega = 10t$, as shown in Fig.~\ref{fig:PT_vs_DMRG}. Note that here we use $\lambda=0.125$ in the anti-adiabatic approximation and $\lambda=0.163$ for the DMRG results. This choice is to partially correct for the 2nd order perturbation theory's 
underestimation of the critical $\lambda_c$, where the single polaron ground state undergoes a sharp transition to a non-zero $k$-value \cite{Marchand10}.  This transition occurs at $\lambda_c=0.125$
in the $\Omega \rightarrow \infty$ limit but at a higher $\lambda_c$ 
for any finite $\Omega$ \cite{Marchand10}. (Note that there is a factor of four difference between the definition of $\lambda$ used here vs. in Ref. \cite{Marchand10}.) The DMRG results with $\lambda=0.163$ appear to be very close to this critical value, based on the flatness of the independent two-polaron continuum located at $E_p(k)+E_P(0)$. To mimic a similar situation, we take a lower value of $\lambda$ in the perturbation theory calculations to place the system in a similar proximity to the transition.

\section{Additional results for electron addition spectral function for extended Hubbard-SSH model}
\begin{figure}[h]
    \centering
    \includegraphics[width=\columnwidth]{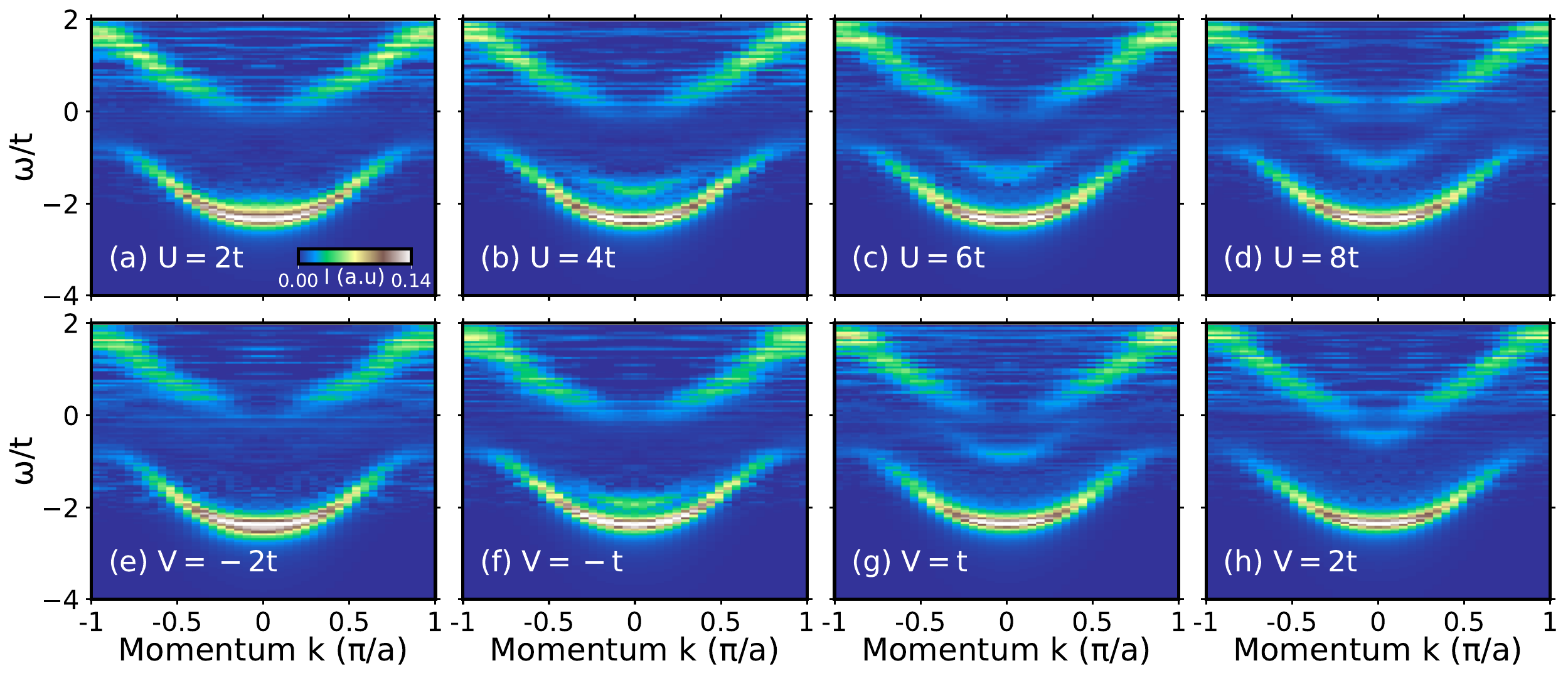}
    \caption{Electron addition spectral function $A^+(k,\omega)$ calculated with DMRG for $L=40$ site chain for the extended-Hubbard-SSH model. All the results are calculated with fixed $g=0.8$, $\Omega=2t$ and $n=0.025$. Panels (a)-(d) show results for fixed $V=0$, and different $U$, whereas, panels (e)-(h) show results for fixed $U=6t$, and different $V$, as indicated in each panel. }
    \label{fig:Akw_U_V_Supp}
\end{figure}

Figure~\ref{fig:Akw_U_V_Supp} shows the momentum-resolved electron addition spectra for several of the cases shown in Fig. 3b and 3e of the main text. 

\section{Additional results for electron addition spectral functions for Hubbard-Holstein model}
\begin{figure}[h]
    \centering
    \includegraphics[width=\columnwidth]{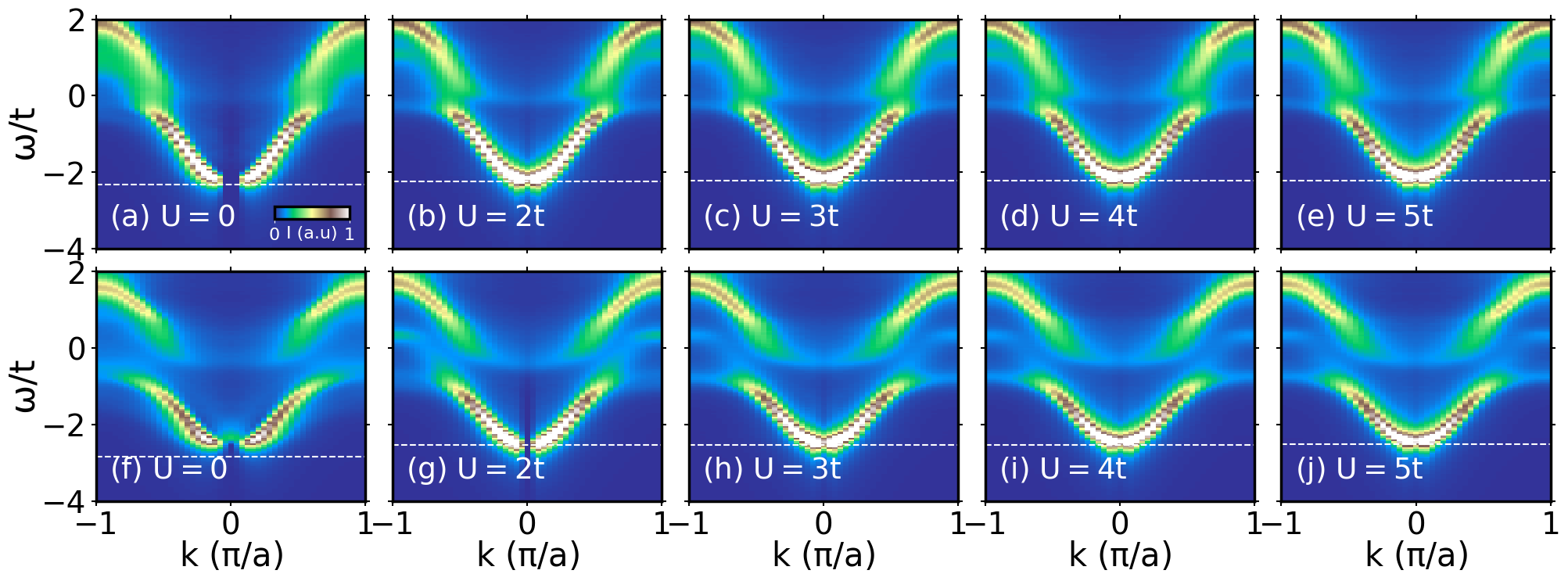}
    \caption{Electron addition spectral function $A^+(k,\omega)$ calculated with DMRG for $L=48$ site chain for the Hubbard-Holstein model. Panels (a-e) show results for fixed $g=1.0$, $\Omega=2t$, $n=0.083$, with varying $U$ as indicated, and panels (f-j) show similar results for $g=1.414$. }
    \label{fig:Akw_HH}
\end{figure}

Figure~\ref{fig:Akw_HH} shows the momentum-resolved electron addition spectral function $A^+(k,\omega)$ for the Hubbard-Holstein model calculated with DMRG for an $L=48$ site chain with $\Omega=2t$, $n=0.083$. Fig.~\ref{fig:Akw_HH}(a-e) show results for fixed $g=1.0$ and varying $U$ as indicated in each panel, while Fig.~\ref{fig:Akw_HH}(f-j) show similar results for $g=1.414$. In all values of $e$-ph couplings and $U$, we do not observe the resonance in the Hubbard-Holstein model, as is observed in the extended-Hubbard-SSH model [Fig.~\ref{fig:Akw_U_V_Supp}]. 

\bibliography{references.bib}